\documentclass[twocolumn]{aastex631}
\usepackage{amsmath}

\begin{document}

\title{{Accretion and Recovery in Giant Eruptions of Massive Stars}}

\shorttitle{Accretion in Massive Stars}

\shortauthors{B. Mukhija and A. Kashi}

\author[0009-0007-1450-6490]{Bhawna Mukhija}
\affiliation{Department of Physics, Ariel University, Ariel, 4070000, Israel}
\email{bhawnam@ariel.ac.il}

\author[0000-0002-7840-0181]{Amit Kashi}
\email{kashi@ariel.ac.il}
\affiliation{Department of Physics, Ariel University, Ariel, 4070000, Israel}
\affiliation{Astrophysics, Geophysics, and Space Science (AGASS) Center, Ariel University, Ariel, 4070000, Israel}

\begin{abstract}
Giant Eruptions (GEs) are episodic high-rate mass loss events that massive stars experience in the late stage of evolutions before exploding as a core-collapse supernova. If it occurs in a binary system, the companion star can accrete part of the mass.
We use numerical simulations to analyze how the companion responds to accretion and how its structure and evolution are altered. We run a grid of massive stars with masses from $20~\rm M_{\odot}$ to $60~\rm M_{\odot}$, and accretion rates from $\rm 10^{-4}$ to $\rm 0.1 ~M_{\odot}~\rm yr^{-1}$, over a duration of $20$ yrs.
For accretion rates $\rm \lesssim 0.01 ~M_{\odot}~\rm yr^{-1}$ the star remains on the hotter side of the HR diagram with a minor increase in luminosity without expanding, as the accretion timescale exceeds the thermal time scale by a larger factor.
Mass loss through stellar winds leads to a minor drop in luminosity shortly after the accretion phase as the star enters the recovery phase. 
For $\rm \gtrsim 0.01 ~M_{\odot}~\rm yr^{-1}$ the companion star experiences a sudden increase in luminosity by about one order of magnitude, inflates, and cools. Under the accreted gas layer the star retains its structure and continues to eject radiation-driven wind during the recovery phase, namely the time it takes to regain equilibrium. Eventually, the accreted material mixes with the inner layers of the star, and the star continues to evolve as a more massive star.
\end{abstract}


\section{Introduction} \label{sec:intro}
\label{1}

Giant Eruptions (GEs) represent transient, high-luminosity outbursts occurring in Luminous Blue Variables (LBVs), typically involving extreme mass-loss episodes and marked variations in spectral and photometric properties. During a GE, the LBV may eject several solar masses of material, particularly in the most extreme cases such in the 1837-1858 eruption of $\eta$ Car \citep[e.g.,][]{1997ARA&A..35....1D, 1999PASP..111.1124H, 2012Natur.486E...1D, 2012ASSL..384....1H}. P Cyg experienced an eruptive behavior in the seventeenth century, leaving a complex circumstellar structure observable today \citep[e.g.,][]{1988IrAJ...18..163D,1994PASP..106.1025H, 2012ASSL..384...43D, 2018NewA...65...29M}.
The expelled gas can significantly reshape the circumstellar environment and contribute to long-term evolutionary deviation from single-star evolution.

GEs are not only critical to the mass-loss history of LBVs, but may also serve as precursors to supernova impostor phenomena, or, in some cases, to terminal core-collapse supernovae. Theoretical modeling, including hydrodynamic simulations and radiative transfer studies \citep{2016ApJ...817...66K}, suggests that the post-eruption recovery phase may span several decades to centuries, during which the star may remain in a state of thermal and structural disequilibrium, often with sustained elevated mass-loss rates driven by radial pulsations or super-Eddington winds.

Massive stars are commonly born in binary or higher-multiplicity systems \citep[e.g.,][]{1998A&ARv...9...63V,2009AJ....137.3358M, 2012Sci...337..444S, 2013ARA&A..51..269D, 2017A&A...598A..84A}, many of which interact with each other and may experience mass exchange or even merge with their companion before completing their evolution \citep[e.g.][]{2008MNRAS.389..925T, 2020ApJ...901...44W, 2024A&A...686A..45S, 2024AzAJ...19b..73M}.
The episodic nature and large-scale circumstellar shaping of GEs underscore their significance in massive stellar evolution, particularly in binary systems where interactions can enhance or even trigger such instabilities \citep{2016ApJ...817...66K, 2021ApJ...914...47K,2022MNRAS.516.3193K}.
Their study informs not only the late-stage evolution of massive stars but also the origin of nebular morphologies and chemical enrichment in the surrounding interstellar medium. Recent work by \citet{2023ApJ...958..138W} examines the impact of mass accretion on the secondary star, and \citet{2024A&A...691A.174S} discuss how angular momentum redistribution affects post-interaction stellar rotation rates. Interactions between components in binary or multiple-star systems, involving mass and angular momentum transfer significantly influence the fundamental parameters and outcomes of both stars \citep[e.g.,][]{2012Sci...337..444S, 2014ApJS..215...15S, 2017ApJS..230...15M, 2023ApJ...958..138W,2024A&A...691A.174S}.

Eruptions of intermediate luminosity optical transients (ILOTs) are a class of stellar events that are characterized by short-lived, but relatively bright optical eruptions \citep[e.g.,][]{2009ApJ...699.1850B, 2012ApJ...746..100S, 2016ApJ...825..105K, 2019ApJ...886...40J, 2021A&A...653A.134B, 2021A&A...646A...1K, 2022MNRAS.511.1330G, 2022Univ....8..493C, 2022A&A...667A...4C, 2023MNRAS.524L..94S}.
These eruptions, also known as Luminous Red Novae (LRNs) or Red Transients, are thought to be caused by the merger of two stars or the interaction between a primary star and a companion. They are often considered intermediate phenomena between novae and supernovae, with luminosities and durations between those two events.
Both ILOTs and GEs are believed to be powered by accretion onto a companion or a merger event, where gravitational energy from these interactions provides the energy for the outbursts \citep[e.g.,][]{2001MNRAS.325..584S,2001A&A...377..672S,2004ApJ...612.1060S, 2010ApJ...709L..11K, 2016MNRAS.462..217S, 2017MNRAS.468.4938K}.

The accretion model for LBV eruptions \citep{2001MNRAS.325..584S, 2004ApJ...612.1060S,Soker_2007,2007MNRAS.378.1609K,2008NewA...13..569K,2016ApJ...825..105K,2017MNRAS.464..775K} suggests that giant LBV eruptions are triggered by periastron passages of a binary companion. For this to happen, the primary star must already be in an unstable state, but it is the secondary star's interaction that amplifies the eruption's intensity.
\citet{2010ApJ...709L..11K} suggested that if the companion star is sufficiently close, it can accrete some of the material ejected by the primary through Bondi-Hoyle-Lyttleton (BHL) accretion \citep{1944MNRAS.104..273B}, and form an accretion disk that launches jets. This accretion model is further supported by the presence of bipolar nebulae around about half of LBVs, which take the form of either hourglass structures or polar caps \citep{2012ASSL..384..171W}.

In binary population synthesis, it is common to assume that the accretor gains a fixed or arbitrary fraction of the mass lost by the donor \citep{2002MNRAS.329..897H}.
A more common approach sometimes used is to adjust accretion based on the initial properties of the accretor when mass transfer begins \citep[e.g.,][]{Bouffanais_2021, Briel_2023, 2024A&A...686A..45S}. In some cases, the companion will accrete a large amount of mass in a short period of time and will have to readjust itself to the new added mass. This topic has not yet received much attention and the effect of the accreted mass on the star is not yet understood.

\citet{2025RAA....25b5010B} found that massive main-sequence (MS) stars can accrete
mass at very high mass accretion rates without expanding much if they lose a significant fraction of
this mass from their outer layers simultaneously with mass accretion. They considered the accretion mechanism to be via an accretion disk that launches powerful jets from its inner zones. These jets remove the outer high-entropy layers of the mass-accreting star. They found that as the star does not expand much, its gravitational potential well stays deep, and the jets are energetic. 
Thus they proposed that these results are relevant to transient events of binary systems powered by accretion and the launching of jets, e.g., ILOTs,
the grazing envelope evolution \citep[e.g.,][]{2020Galax...8...26S, 2023MNRAS.524L..94S}, and the great eruption of $\eta$ Car.
\cite{Scolnic_2025} simulated mass accretion at very high rates onto
massive MS stars combined with mass loss and found that they can accrete up to $10\%$ of their mass without expending much if the mass is being removed by jets, and keep their deep potential well.

\citet{2024ApJ...966L...7L} derived fitting formulae for the maximum radii of rapidly accreting stars, ``hamsters'', whose accretion rates exceed their thermal acceptance. These conditions arise in case B and C mass transfer \citep{1977ARA&A..15..127T}, where the donor’s thermal timescale is shorter than the accretor’s. Above a critical accretion rate, the accretor expands along the Hayashi line, resembling a red giant, before contracting toward the zero-age main sequence (ZAMS) as it becomes luminous enough to radiate away excess energy. They provided semi-empirical expressions for the settling radius and a mass-luminosity relation along the Hayashi line, enabling estimates of mass-transfer efficiency from donor mass-loss rates.

In \cite{MukhijaKshi2024a} we investigated the effects of a GE on a massive star by creating an artificial eruption and examining the changes in the star's stellar characteristics and its evolutionary track on the HR diagram during the eruption phase. We used a high mass loss rate of approximately $\rm 0.15~M_{\odot}~yr^{-1} $ to the star's outer layer over a short period. The results show that the star's luminosity decreased and its temperature increased due to the mass loss process. Since \cite{MukhijaKshi2024a} focused on the mass-losing star in the binary, this paper makes the transition to address the mass-gaining star.

In this paper, we study how a massive star reacts when it accretes mass at very high accretion rates. We simulate a single star assuming it is a companion star in a binary system, accreting material at a very high rate due to the GE of the primary star by the wind accretion process.
The paper is organized as follows:
Section \ref{2} discusses the basic assumptions and modeling method for the grid of massive stars. 
Section \ref{3} presents the analysis of our results for the 30 $\rm M_{\odot}$. We begin by describing the evolutionary tracks of 30 $\rm M_{\odot}$ for all four accretion rates. Next, we discuss the results for accretion rates ($\rm \gtrsim 0.01 ~M_{\odot}~\rm yr^{-1}$) that cause fluctuations in luminosity, followed by an examination of the star's surface properties during the accretion phase. Later on we show the results of the recovery phase of the star. Section \ref{4} presents our conclusions and section \ref{5} a summary of the findings.

\section{The Numerical Simulation}
\label{2}

\begin{figure}
    \centering
\includegraphics[trim={0.12cm 0.0cm 1.1cm 0.2cm},clip,width=1\columnwidth]{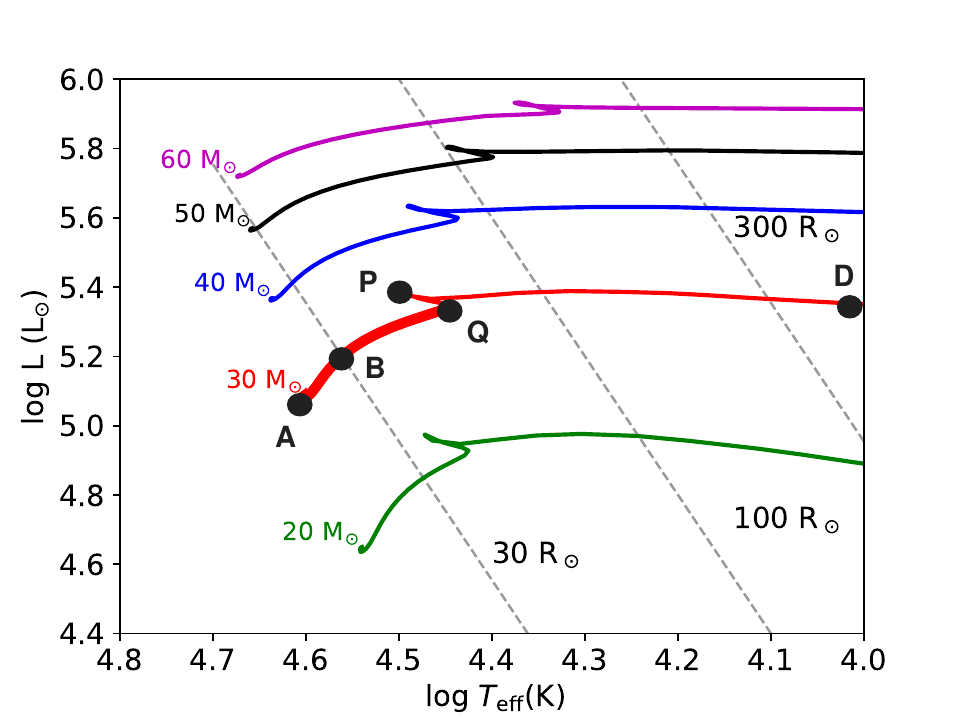}
     \caption{Evolutionary tracks obtained using the MESA default. Point A: ZAMS. Point B: Beginning of the accretion phase. Point Q to P: Terminal age MS phase (TAMS; Henyey hook). Point D: End of calculation.
     }
    \label{HR0}
 \end{figure}

Using the \textsc{mesa} stellar evolution code \citep[Modules for Experiments in Stellar Astrophysics -- version r23.05.1;][]{2011ApJS..192....3P, 2013ApJS..208....4P, 2015ApJS..220...15P, 2018ApJS..234...34P, 2019ApJS..243...10P}, we created a grid of models of massive stars with masses ranging from 20 to 60 $\rm M_{\odot}$. We examine four accretion rates for each mass, ranging from $\rm 10^{-4}$ to $\rm 0.1~M_{\odot}~\rm yr^{-1}$ over $20$ yrs during the accretion phase.

The accretion rate onto the companion during the Great Eruption of $\eta$ Car was very high and could reach up to $\simeq 0.3 ~M_{\odot}~\rm yr^{-1}$ over its $\simeq 20$ years of eruption \citep{2010ApJ...723..602K}.
The eruptions of P Cygni during the nineteenth century also extended over yrs to decades and were conjectured to have transferred mass to a companion star at an average rate of $\simeq 0.02 ~M_{\odot}~\rm yr^{-1}$ intermittently over $\simeq 40$ yrs \citep{2010MNRAS.405.1924K}.
For SN~2009ip the rates might have reached higher values, but the duration was shorter \citep{2013MNRAS.436.2484K}.
As the above systems are rare, there are likely GEs where the mass loss rate from the LBV is smaller and the orbital separations are larger, so we consider a range of $\rm 10^{-4}$ to $\rm 0.1 ~M_{\odot}~\rm yr^{-1}$ over $20$ yrs as a reasonable range of parameters for a first numerical experiment of its kind, although other values should be tested later as well.

We assume that the star is the companion star in a binary system with a more massive star being the primary. We further assume that the primary undergoes a GE, during which mass is being lost from its surface at a high rate. The close orbit is assumed to result in accretion at a high rate of primary material onto the companion during the event of an eruption \citep{2010ApJ...709L..11K}. Our simulation does not treat the primary but focuses on the accreting companion as a single star, and follows its evolution under the influence of accretion. Our calculations assume spherical symmetric accretion. We evolve the companion star from the pre-main sequence (pre-MS) stage by using the default option in \textsc{mesa}. We consider the presence of strong stellar winds by using the \textsc{mesa} `Dutch’ wind prescription with a scaling factor of 0.5 \citep{2020MNRAS.495..249Z}.

The accretion phase begins at point B, an evolutionary stage beyond the ZAMS, where $\log T_{\rm eff}  \simeq 4.58 \,\rm K$ (see Figure \ref{HR0}). This temperature is approximately that of the temperature of the companion of $\eta$ Car \citep{2010ApJ...710..729M}. Figure \ref{HR0} shows the evolutionary tracks of the accreting star starting from the ZAMS phase. During the accretion phase, we remove the Dutch wind prescription. Table \ref{T1} presents the stellar parameters of the initial stage of the accretion phase; i.e., point B for our $\rm  30~M_{\odot}$ star. At this point, we ran four simulations with four different $ \dot{M}_{\rm acc} $ ranging from $\rm 10^{-4}$ to $\rm 0.1 ~M_{\odot}~\rm yr^{-1}$, and evolve the star with these mass accretion rates for 20 $ \rm yrs$. The accretion occurs in the outer layers of the star on a timestep ($ \simeq 500~\rm s$, i.e., accretion timestep) that is shorter than the thermal timescale of the star \citep{1982ApJ...253..798N, 2004ApJ...600..390T, 2011ApJS..192....3P, 2013ApJS..208....4P}. The choice of accretion timestep allows us to accurately capture the impact of accretion on the star and observe its effects in detail. After the 20 $ \rm yrs$ of accretion, we proceed to evolve the star further (  $ \rm \simeq 4~Myrs$) and examine its recovery.
We then trace how the recovery phase and evolutionary tracks of the star depend on the amount of mass and energy added into the star's outer layer over the 20 $\rm yrs$ and how the star responds to the accumulation of surface material due to accretion.

 \begin{table}
    \centering
    \begin{tabular}{l c}
    \hline 
    \hline
     $M_{ \rm ZAMS} ~( \rm M_{\odot})$ & 30  \\
     \hline
     $ \rm star~age~ (Myrs) $ & 1.1594326\\
     $M ~( \rm M_{\odot})$ & 29.85  \\ 
     $\log T_{\rm eff} ~(\rm K) $&  4.58 \\
      $\log R ~(\rm R_{\odot}) $ &  0.910\\
     $\log g ~(\rm cm~s^{-2})$ & 4.09\\
     $\log \rho ~(\rm g~cm^{-3})$ & -9.17 \\
     $\log L ~(\rm L_{\odot})$ & 5.12\\

   \hline \hline
    \end{tabular}
    \caption{Stellar parameters correspond to the initial stage of our star (point B in Figure \ref{HR0}) when the experiment starts. The rows give the star age, the mass of the star ($ M$), surface temperature ($ T_{\rm eff}$), surface gravity ($ g $), density ($ \rho$), surface luminosity ($ L$), and mass loss via stellar winds ($ \dot{M}$), respectively.}
    \label{T1}
\end{table}

To set the initial conditions of the star we evolve a non-rotating star with metallicity $Z = 0.02$. In \textsc{mesa}, the parameter for the initial mass fraction is defined as $ Y = Y_{\rm prim} + (\Delta  Y /\Delta Z) / Z$, where $ Y_{\rm prim}$ represents the He abundance,  $ Z$ is the metallicity and $ \Delta Y$ and $ \Delta Z$ represent the changes in them during the evolution. The values we use in our model are $ Y_{\rm prim}=0.24$ and $(\Delta  Y / \Delta  Z) / Z =2$, consistent with the default values in \textsc{mesa} \citep{2009yCat..72980525P}. To model convection mixing, we employ the Mixing Length Theory \citep[MLT;][]{1958ZA.....46..108B,1965ApJ...142..841H}, with mixing length parameter, $\alpha_{\rm MLT}$  = 1.5. Semi-convection is modeled with an efficiency parameter $\rm \alpha_{sc}= 0.01$.
Following \citet{1980A&A....91..175K}, we select the efficiency parameter for the thermohaline mixing $\rm \alpha_{th} $ to be 2.0. We use \citet{2000A&A...360..952H} diffusive technique to represent the convective overshooting, using $f_{1}$ = 0.005 and $f_{0}$ = 0.001 for each convective core and shell. We use an opacity table of type II \textsc{OPAL} which allows for time-dependent variation in the metal abundance in our model \citep{1996ApJ...464..943I}.
Simulating accretion in \textsc{mesa} requires a special approach that handles the interaction of the arriving material with the outer layers of the star. In each step, gravitational energy is added to the gas in the outermost cell.
The details of how accretion is treated are given in appendix \ref{treatment}.

\section{Results}
\label{3}
\subsection{Accretion Simulation}

Our numerical experiment uses four different accretion rates and five values of mass of companion stars companion stars to simulate accretion, with all of our grid consisting of simulations. For every mass accretion rate we start the accretion at the same point B (Figure \ref{1}). We will focus here on the $\rm 30~M_{\odot}$ accreting companion star, which we define as our fiducial simulation. Figure \ref{H_all} displays the evolutionary tracks of a $\rm 30~M_{\odot}$ star under four distinct $\dot{M}_{\rm acc}$, and it is evident from the evolutionary tracks that the response of the star differs for lower and higher $\dot{M}_{\rm acc}$. Similar to the process of accretion, Figure \ref{H_all} also illustrates the recovery phase, highlighting the distinct variations in recovery for lower and higher $\dot{M}_{\rm acc}$.

\subsubsection{Response to low accretion rates}

First, it is important to note that we refer to the accretion rates as `low', but it is a relative term. For most stellar binaries, even massive ones, the rates considered here, $10^{-4}$--$10^{-3}$~$\rm M_{\odot}~\rm yr^{-1}$, would be very high. For example, the massive MS binary HD 166734 \citep{2020MNRAS.492.5261K} and the B[e] supergiant binary system GG Car \citep{2023MNRAS.523.5876K} both accrete at rates of $10^{-7}$--$10^{-6} ~\rm M_{\odot}~\rm yr^{-1}$, and $\eta$ Car accretes at few $\times 10^{-6} ~\rm M_{\odot}~\rm yr^{-1}$ \cite{2017MNRAS.464..775K}.
The rates we study here are orders of magnitude greater than the typical quiescent accretion rates. However, during GEs, even higher accretion rates can occur \citep[e.g.,][]{2010ApJ...723..602K, 2013MNRAS.436.2484K,2025PASP..137c4201S} so we will refer to this range as low accretion rates for the context of this study.

\begin{figure*}
  \centering
  \begin{tabular}{c @{\qquad} c }
    \includegraphics[width=.5\linewidth]{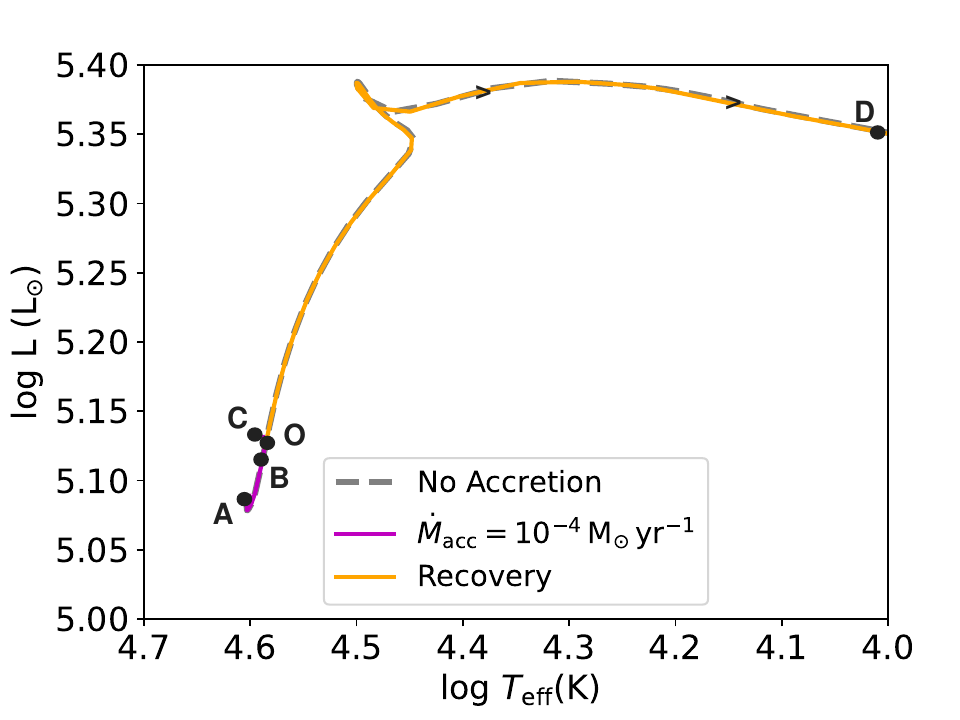}
    &
     \hspace{-1.0cm} 
    \includegraphics[width=.50\linewidth]{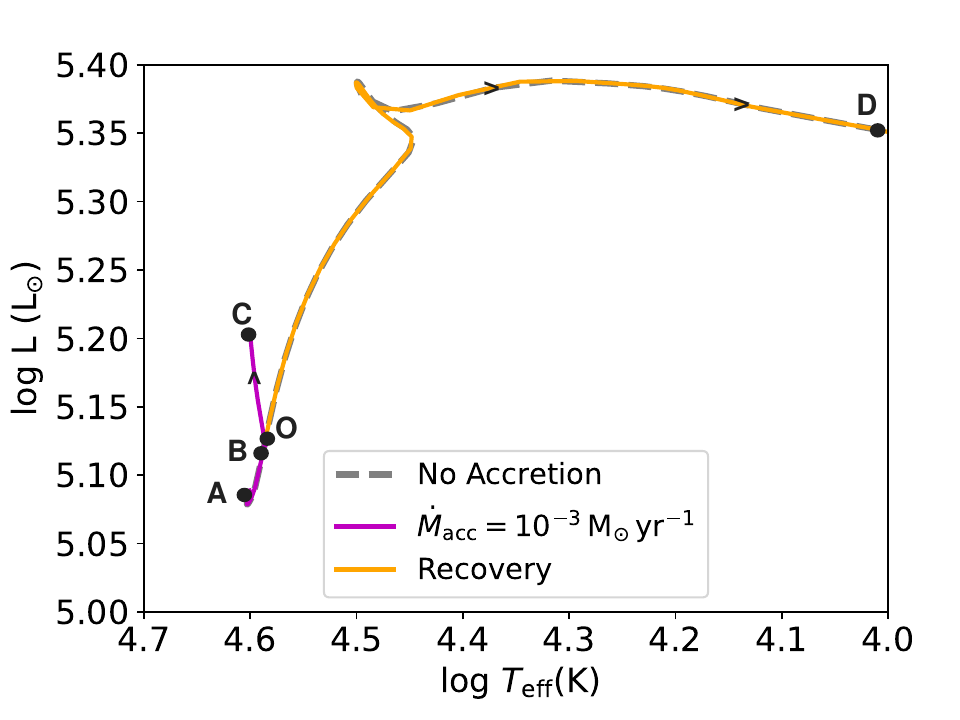}
    \\
    \small (a) & (b)
  \end{tabular}
   \begin{tabular}{c @{\qquad} c }
    \includegraphics[width=.5\linewidth]{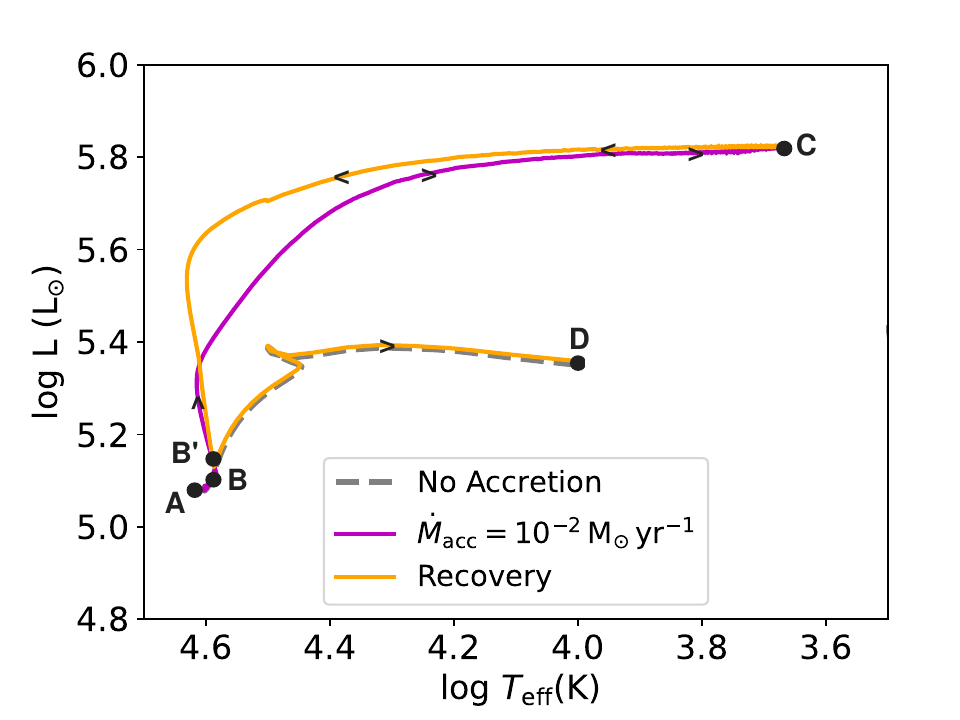}
    &
     \hspace{-1.0cm} 
    \includegraphics[width=.5\linewidth]{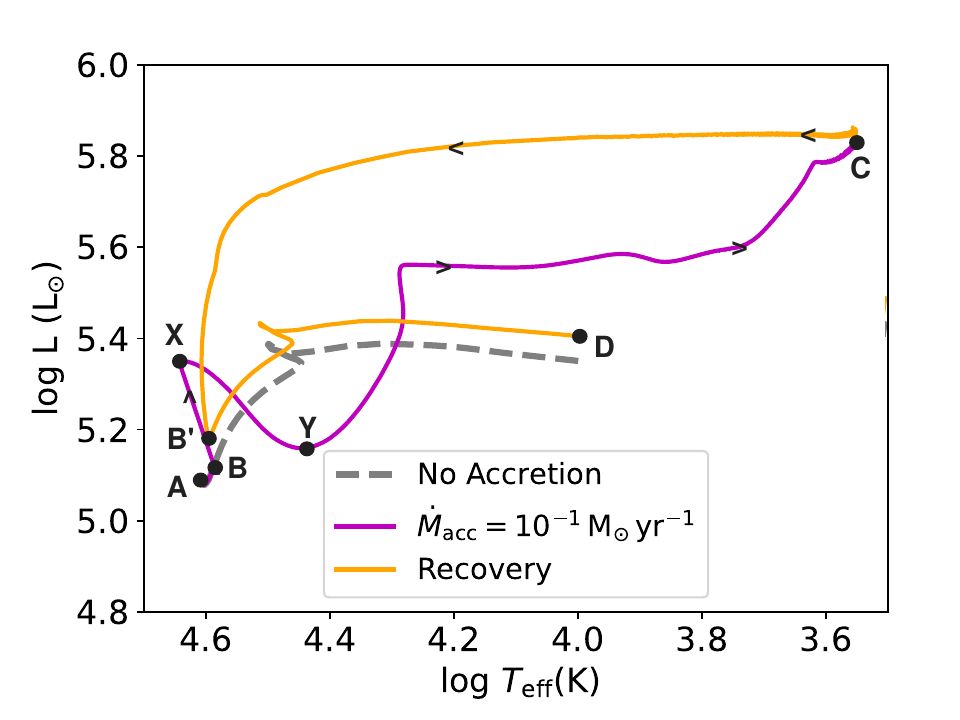}
    \\
    \small (c) & (d)
    
  \end{tabular} 
  \caption{Evolutionary tracks of a $\rm 30~M_{\odot}$ star for lower and higher accretion rates obtained using \textsc{mesa}. The dashed gray background track shows the evolutionary track of the star without the accretion.
  Panel (a) represents the evolutionary track for an accretion rate of $\rm 10^{-4}~M_{\odot}~\rm yr^{-1}$, and panel (b) represents the evolutionary track for an accretion rate of $\rm 10^{-3}~M_{\odot}~\rm yr^{-1}$. 
  In both panels, the tracks from points B to C show the accretion phase, while the tracks from points C to O and O to D represent the recovery phase of the star.  Panel (c) represents the evolutionary track for an \textit{accretion rate $\rm 10^{-2}~M_{\odot}~\rm yr^{-1}$}, and panel (d) represents the evolutionary track for an accretion rate $\rm 10^{-1}~M_{\odot}~\rm yr^{-1}$. 
  The track from point B to C in panel (c) shows the accretion phase, and C to D (from point C to B', and B' to D) shows the recovery phase for an accretion rate $\rm 10^{-2}~M_{\odot}~\rm yr^{-1}$. 
  In panel (d), the track from point B to C (from B to X, X to Y, and Y to C) shows the accretion phase, and C to D (from C to B', and B' to D) represents the recovery phase for an accretion rate $\rm 10^{-1}~M_{\odot}~\rm yr^{-1}$.}
  \label{H_all}
  \end{figure*}

Figure \ref{H_all} panel (a) shows the evolutionary track of $ \rm 30~M_{\odot}$ star with an accretion rate of $\rm 10^{-4}$~$\rm  M_{\odot}~\rm yr^{-1}$.  Here the track from point A to B represents the MS phase, the track from point B to C shows the accretion phase, and from C through O to D represents the star's recovery phase. In the accretion scenario, as material accumulates in the star's outer layers, both its luminosity and temperature increase. 
During this phase, the star's accretion timescale surpasses its thermal timescale, allowing the star to continuously adjust and maintain thermal equilibrium throughout the accretion process. As a result, the star remains on the hot side of the HR diagram during the 20 yrs accretion phase.
After 20 yrs, the star enters the recovery phase, shown by the track from point C through O to D. Due to its high temperature at point C, the star undergoes stellar wind mass loss, as we have applied the Dutch wind prescription during the recovery phase. This mass loss leads to a drop in luminosity, resulting in a vertical descent in the evolutionary track from C to O. Between points C and O, the star loses $ \Delta M = 1.5842 \times 10^{-5}~\rm M_{\odot} $  in $ \Delta t = 10^{3}~\rm yrs$. Following this, the star experiences radial expansion and shifts toward the cooler side of the HR diagram during the recovery phase. The star enters the helium-burning phase and reaches it at point D.

Similarly, Figure \ref{H_all} panel (b) shows the evolutionary track for an accretion rate of $\rm 10^{-3}$~$\rm M_{\odot}~\rm yr^{-1}$.
The evolutionary track follows a similar pattern as in Figure \ref{H_all}, panel (a): from A to B represents the MS phase, B to C the accretion phase, and C through O to D the recovery phase. While at the end of the accretion phase (point C), the star has a higher luminosity and temperature than at $\rm 10^{-4}$~ $\rm M_{\odot}~\rm yr^{-1}$ due to the increased accreted material. The star then enters the recovery phase, shown by the track from point C through O to D. Here also, the track from C to O shows a drop in luminosity due to stellar wind mass loss, similar to Figure \ref{H_all}, panel (a). Between points C and O, the star loses $ \Delta M = 0.358~\rm M_{\odot} $ in $ \Delta t = 0.263~\rm M yrs$. After reaching point O, the star expands toward the cooler side of the HR diagram and reaches at point D during this recovery phase. For these lower accretion rates the estimated time from point C to D is $\simeq 4.06$ Myrs.

\subsubsection{Response to high accretion rates}
Figure \ref{H_all}, panel (c), shows the evolutionary track for an accretion rate of $10^{-2}~\rm M_{\odot }~\rm yr^{-1}$.
Compared to lower accretion rates, the star behaves differently here, it inflates and moves towards the cooler side of the HR diagram during the accretion phase. It experiences a rapid increase in luminosity during the first $ \simeq 413 ~\rm days $ of accretion, after which it maintains a more constant luminosity over the remaining time until 20 yrs are reached. This suggests that surface luminosity is driven up almost immediately and adiabatically towards the accretion luminosity as mentioned in \citet{2024ApJ...966L...7L}. Based on these outcomes we can conclude that (i) after adding the mass, the luminosity rises steeply due to accretion before the star expands appreciably (ii) most of the expansion takes place before a substantial amount of the mass has been accreted. Thus at the high accretion rates, as mass is added to a low-density envelope outside the initial stellar radius, accreting MS stars expand significantly. The star cannot fully adjust to the incoming mass at its surface, leading it to exit thermal equilibrium and inflate towards the cooler side of the HR diagram

\begin{figure}
    \centering
\includegraphics[trim={0.7cm 0.7cm 1.6cm 2.4cm},clip,width=1\columnwidth]{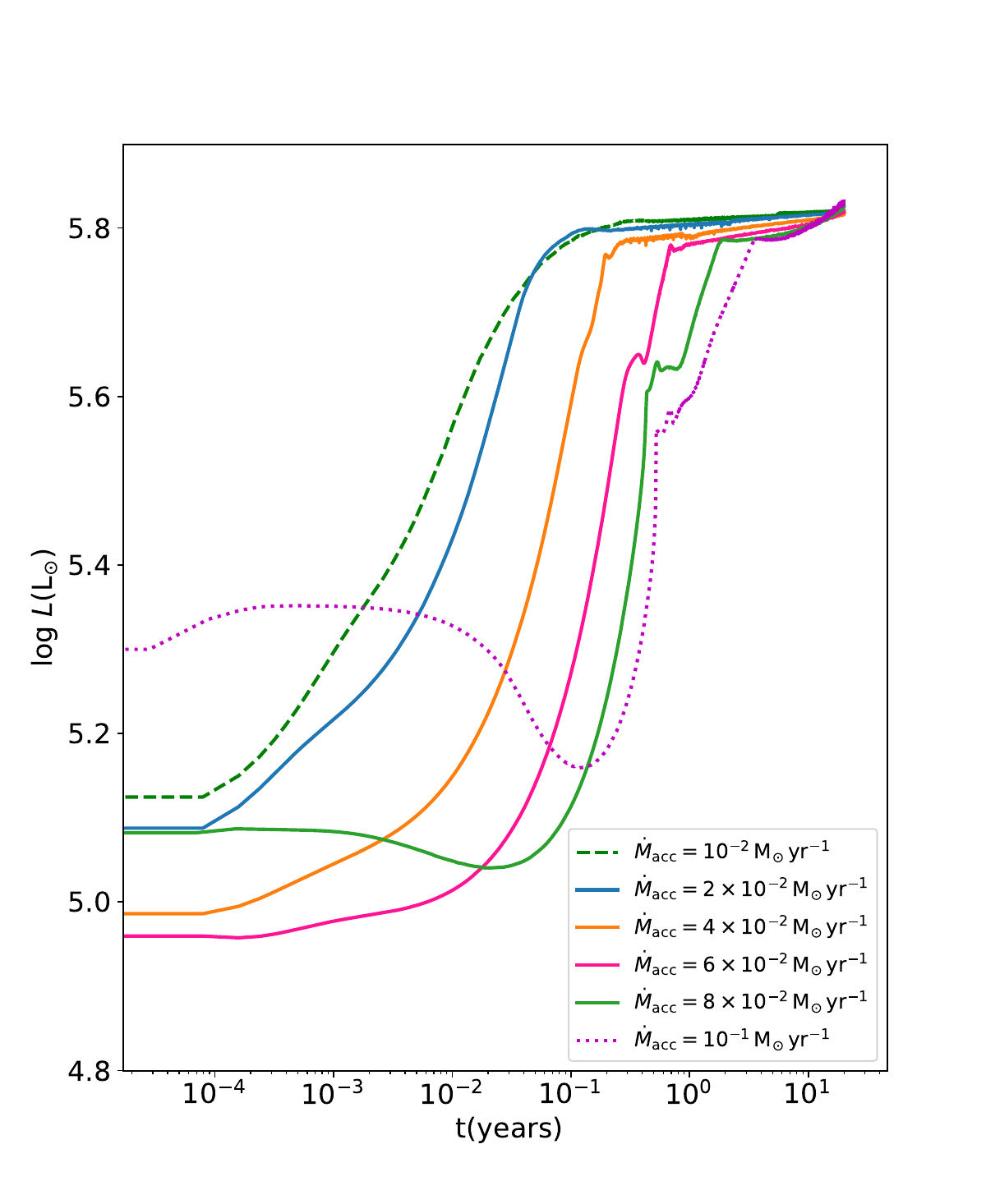}
     \caption{The luminosity variation with time during the accretion phase for 6 tracks corresponding to the 6 distinct accretion rates (see legend) over the accretion phase lasting 20 $ \rm yrs$. The green line ($\rm 8 \times 10^{-2}~M_{\odot}~\rm yr^{-1}$) and dotted magenta line ($\rm 10^{-1}~M_{\odot}~\rm yr^{-1}$) show the luminosity fluctuation due to compression.}
     \label{L}
 \end{figure}

After point C, the star enters the recovery phase and shifts towards the hotter side of the HR diagram. The track from  C to B', and B' to D then represents this phase. As the accretor becomes more luminous at point C, compared to point B, it more quickly transports and radiates away the excess gravitational energy from the added material. This causes the star to contract and evolve towards the ZAMS back after reaching its maximum radius (i.e., point C). At the beginning of the recovery phase, the star’s surface temperature is low ($\log T_{\rm eff} = 3.66~\rm K$), and stellar wind mass loss is minimal. However, as the star contracts more toward the hotter side of the HR diagram, mass loss due to stellar winds begins. During the track from point C to B', the outer layers have a lower density than at point C, due to allowing material to be blown away in this phase. The star then enters the helium-burning phase and reaches to point D, after passing the TAMS, while remaining in thermal equilibrium. The evolutionary track passes close to point B, where the accretion process originally began, now marked as point B' in Figure \ref{H_all} panel (a). 

\begin{figure*}
  \centering
  \begin{tabular}{c @{\qquad} c }
    \includegraphics[width=.5\linewidth]{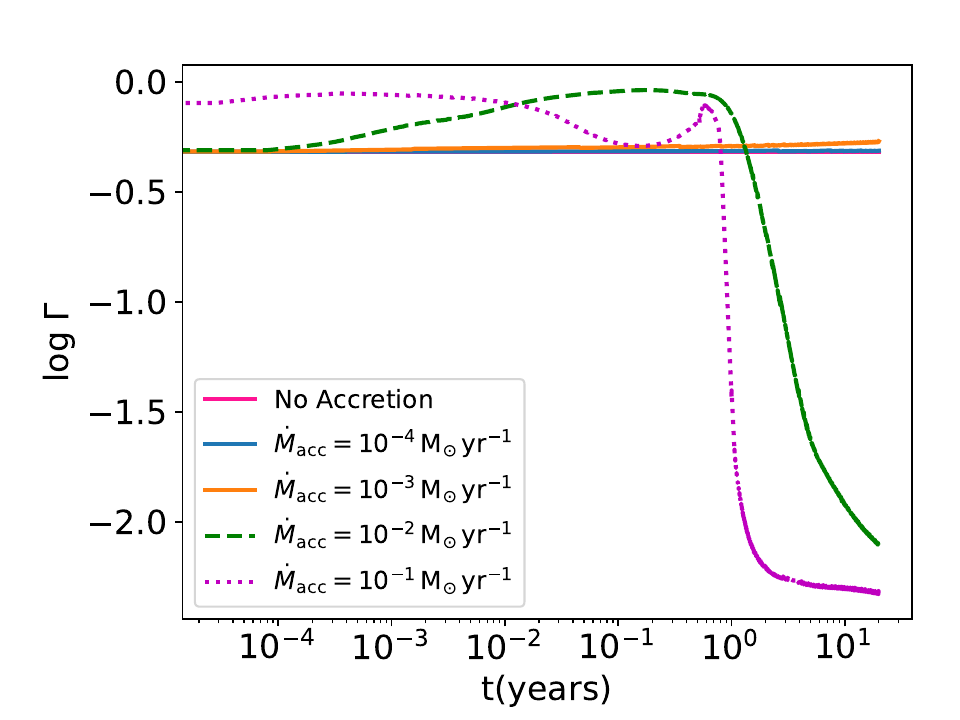}
    &
     \hspace{-1cm} 
    \includegraphics[width=.5\linewidth]{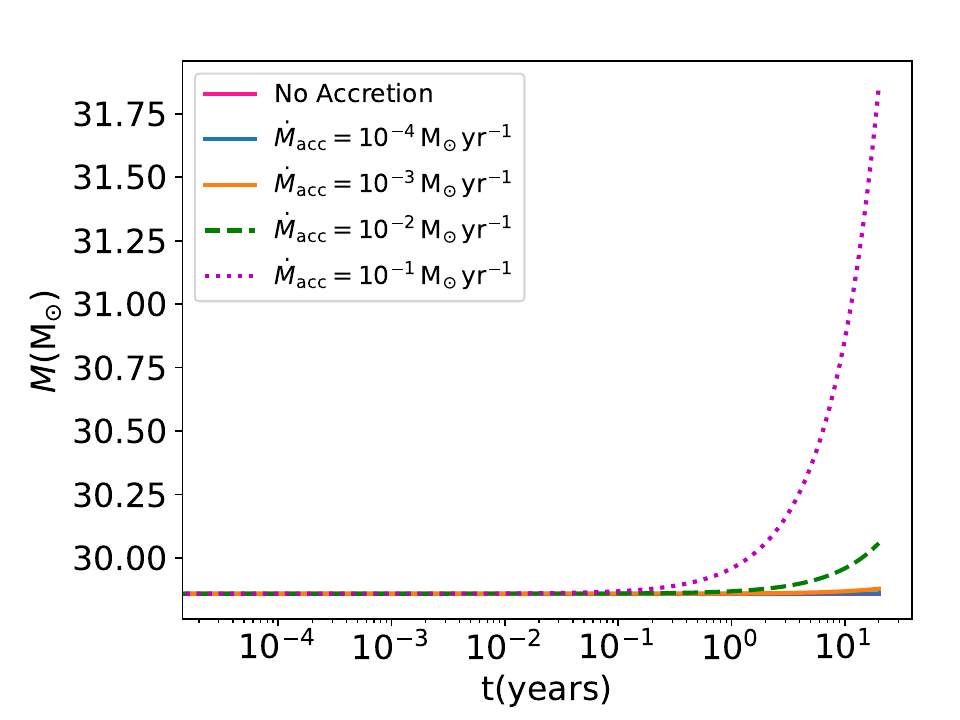}
    \\
    \small (a) & (b)
  \end{tabular}
   \begin{tabular}{c @{\qquad} c }
    \includegraphics[width=.5\linewidth]{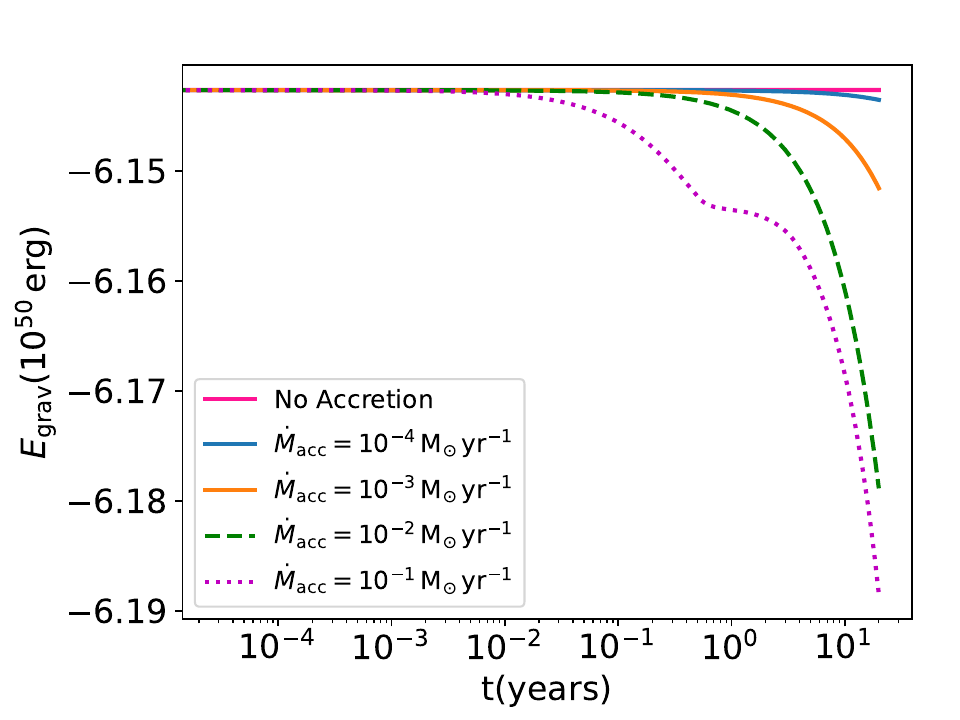}
    
    &
     \hspace{-1cm} 
    \includegraphics[width=.50\linewidth]{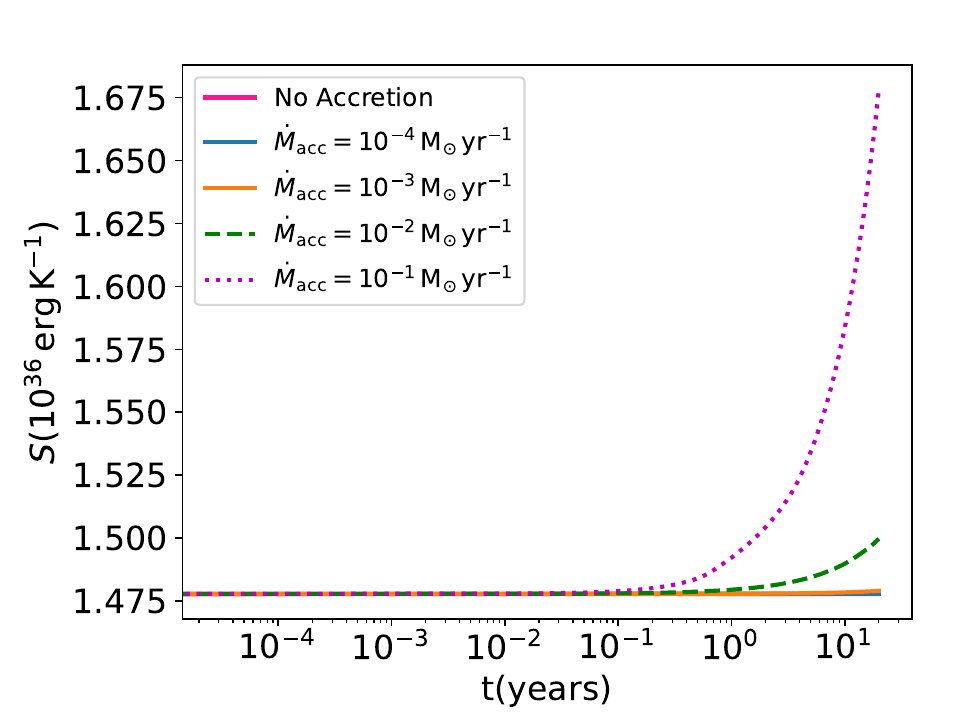}
    \\
    \small (c) & (d)
  \end{tabular}
  
  \caption{Stellar tracks of the  Eddington parameter ($ \Gamma$), stellar mass ($M$), gravitational energy ($E_{\rm grav}$), and total entropy ($S$) following a period of 20 $\rm yrs$ during the accretion phase is depicted in panel (a), panel (b), panel (c), and panel (d) respectively. In panel (a), the $\Gamma$ value represents the strength of the stellar winds responsible for blowing up material from the star's outer layers. However, in the accretion phase, wind mass-loss prescription is not considered, leading to a decrease in the $ \Gamma$ value over 20 $ \rm yrs$. Panel (b) illustrates the varying stellar mass during the accretion phase for different accretion rates. 
In panel (c) and panel (d), the variations in gravitational energy and entropy during the accretion phase are depicted.}
  \label{all}
\end{figure*}
\begin{figure}
\centering
\begin{tabular}{@{}c@{}}
    \includegraphics[scale=0.52]{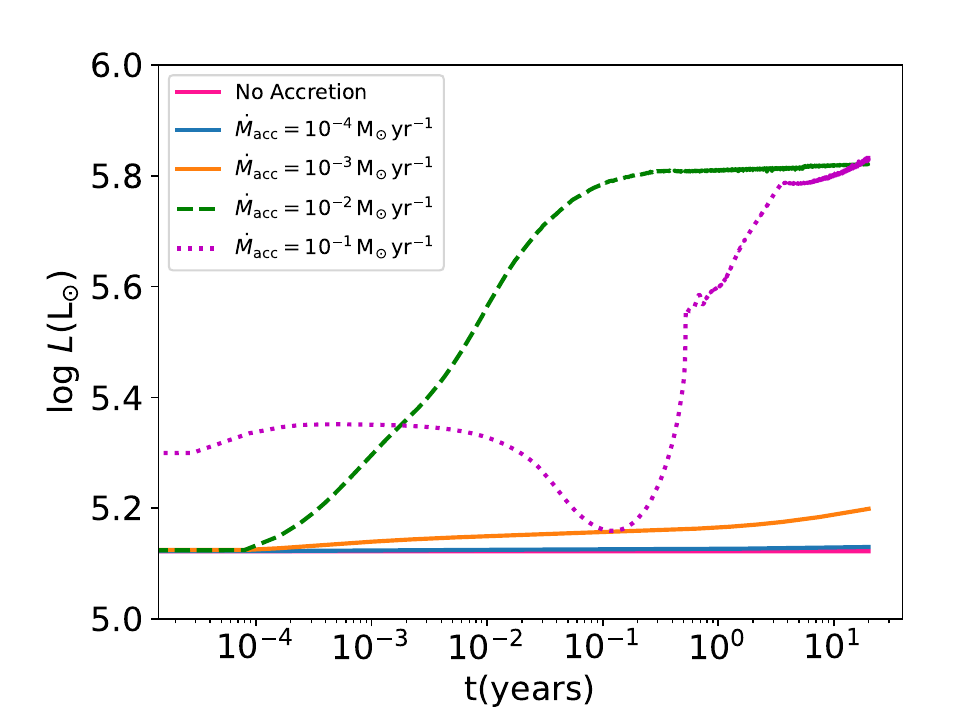} \\
    \small (a) \\
    \includegraphics[scale=0.52]{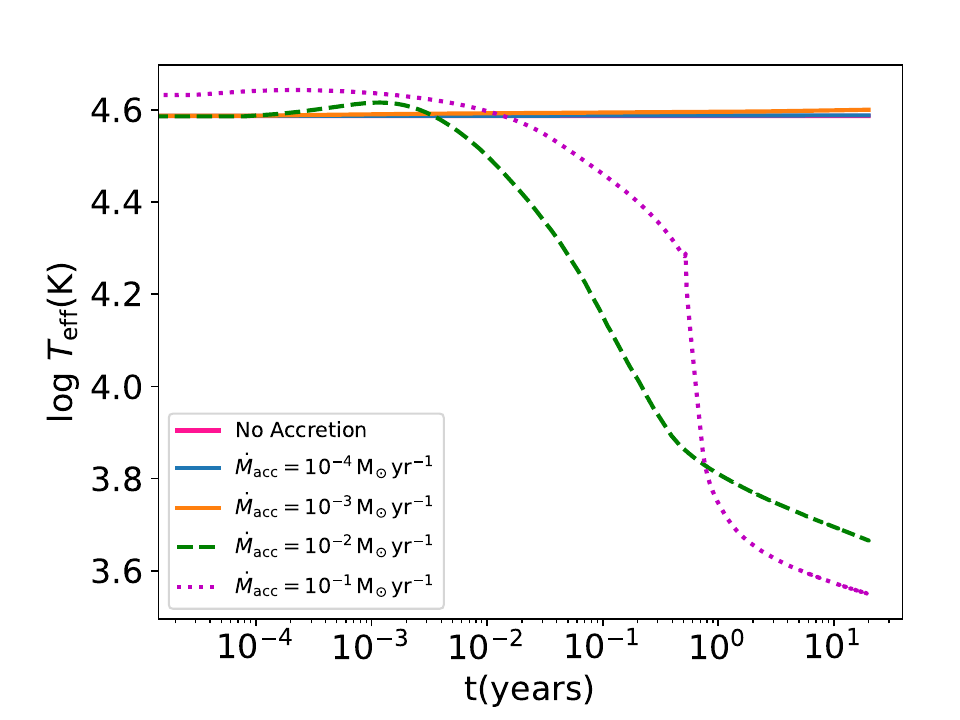} \\
    \small (b) \\
    \includegraphics[scale=0.52]{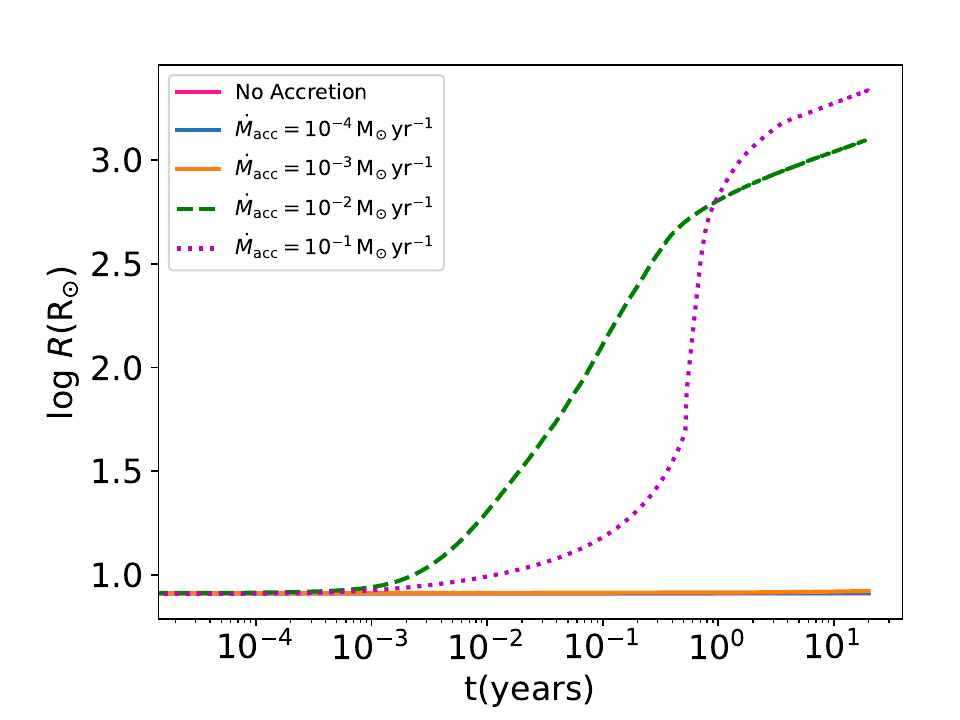} \\
    \small (c)
\end{tabular}
\caption{The variation in luminosity $L$ (panel a),
effective temperature $T_{\rm eff}$ (panel b), and radius $R$ (panel c) of the star during the mass accretion process for four different accretion rates. The differences in these parameters between high and low accretion rates are evident.}
\label{LTR}
\end{figure}
\begin{figure*}
  \centering
  \begin{tabular}{c @{\qquad} c }
    \includegraphics[width=.52\linewidth]{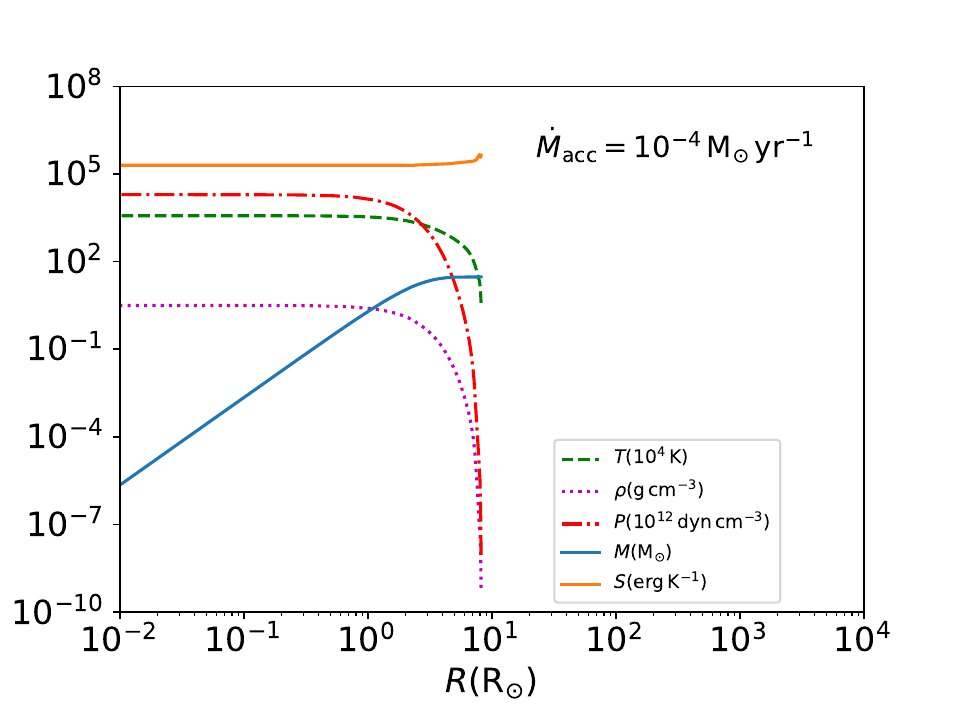}
    &
     \hspace{-1.2cm} 
    \includegraphics[width=.52\linewidth]{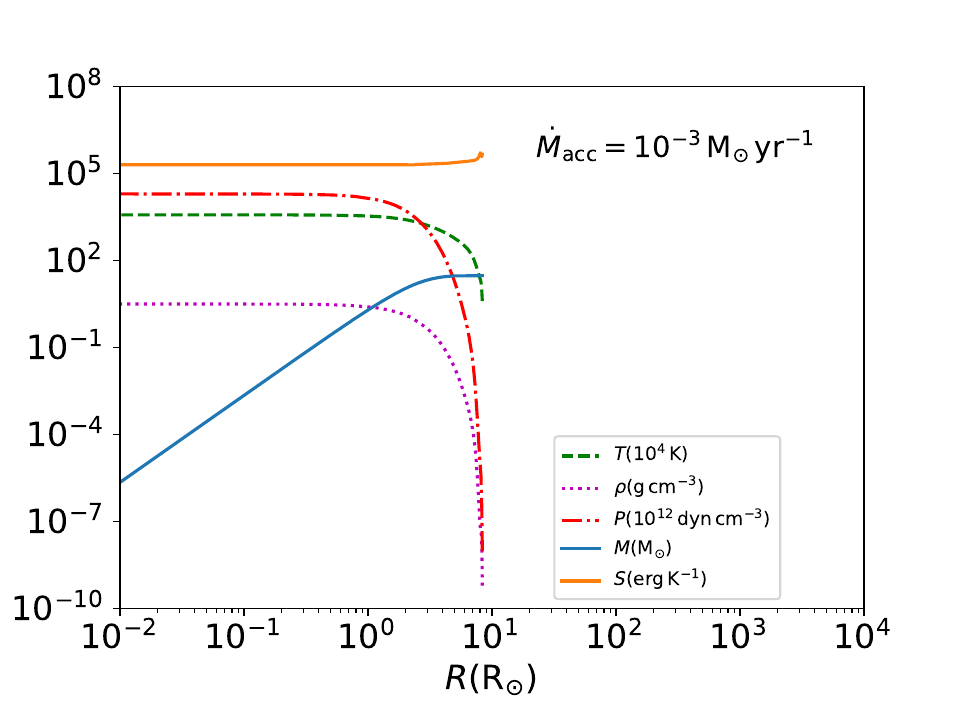}
    \\
    \small (a) & (b)
  \end{tabular}
   \begin{tabular}{c @{\qquad} c }
    \includegraphics[width=.52\linewidth]{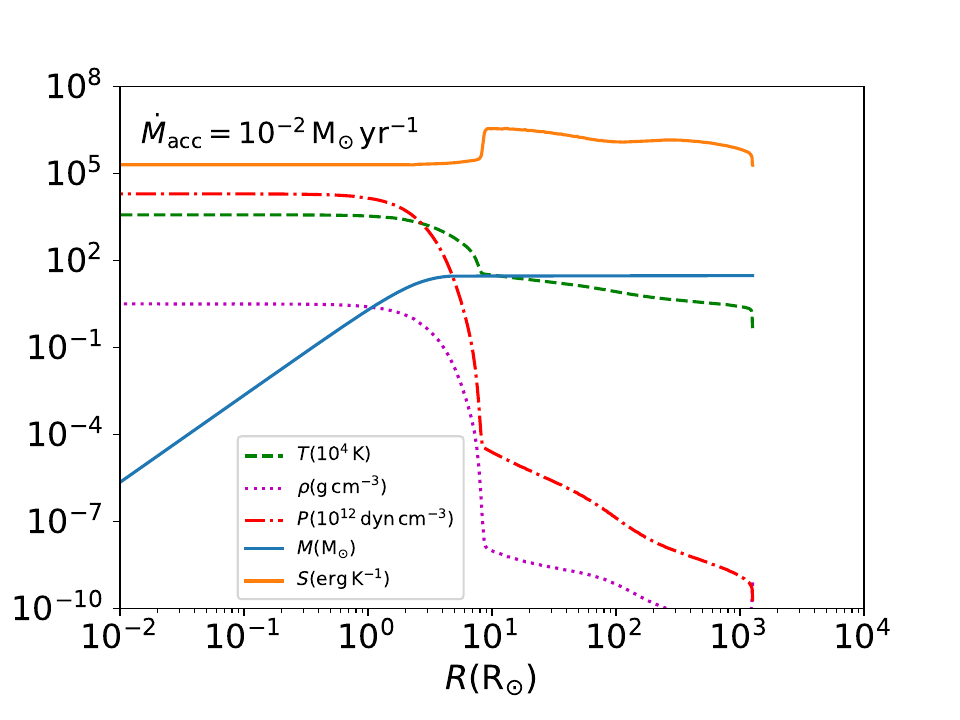}
    
    &
     \hspace{-1.2cm} 
    \includegraphics[width=.52\linewidth]{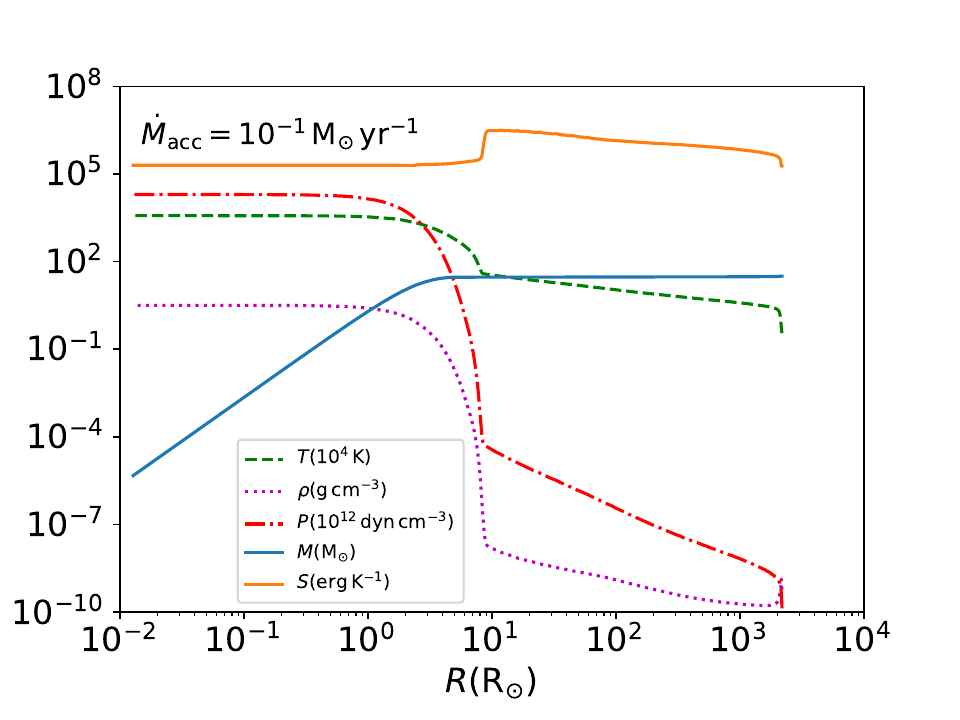}
    \\
    \small (c) & (d)
  \end{tabular}
  \caption{At the final stage of the accretion phase (point C, (see Figure \ref{H_all})), the stellar properties, including temperature, density, pressure, mass, and entropy of a $\rm 30~M_{\odot}$ star for four different accretion rates are depicted above. Panel (a), panel (b), panel (c), and panel (d) correspond to the accretion rates $\rm 10^{-4}$, $\rm 10^{-3}$, $\rm 10^{-2}$, and  $\rm 10^{-1}~M_{\odot}~\rm yr^{-1}$ respectively. Panels (a) and (b) show that for lower accretion rates, after the accretion of material, the radius of the outer region is approximately 10 $\rm R_{\odot}$. In contrast, for higher accretion rates, the star inflates and crosses 2000 $\rm R_{\odot}$. Similarly, other stellar parameters like temperature, density, pressure, and entropy are higher in the outer region for lower accretion rates compared to higher accretion rates.
  The star at this point is not one unit but rather composed of the original star and above which there is a layer that came from the accreted material and is clearly separated in all its properties.}
  \label{pointC}
\end{figure*}
\begin{figure*}
  \centering
  \begin{tabular}{c @{\qquad} c }
    \includegraphics[width=.52\linewidth]{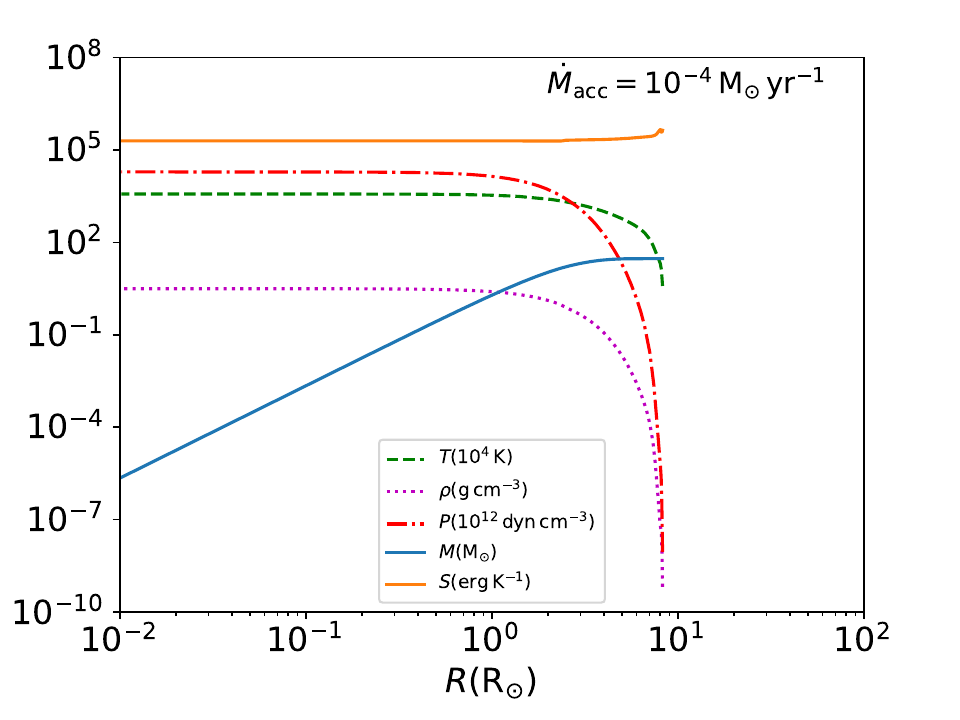}
    &
     \hspace{-1.2cm} 
    \includegraphics[width=.52\linewidth]{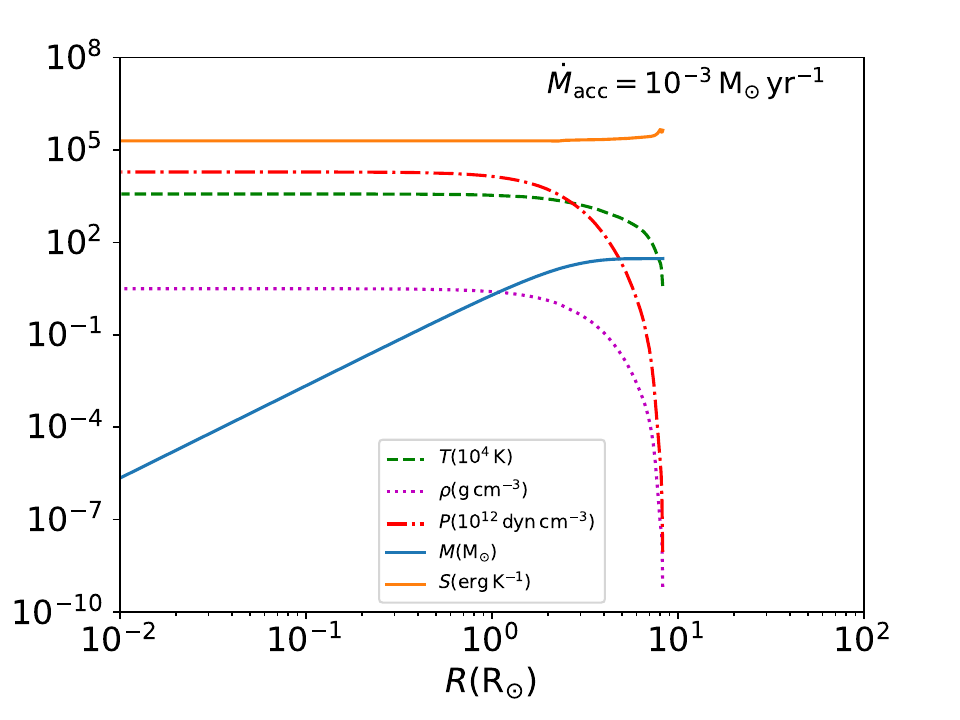}
    \\
    \small (a) & (b)
  \end{tabular}
   \begin{tabular}{c @{\qquad} c }
    \includegraphics[width=.52\linewidth]{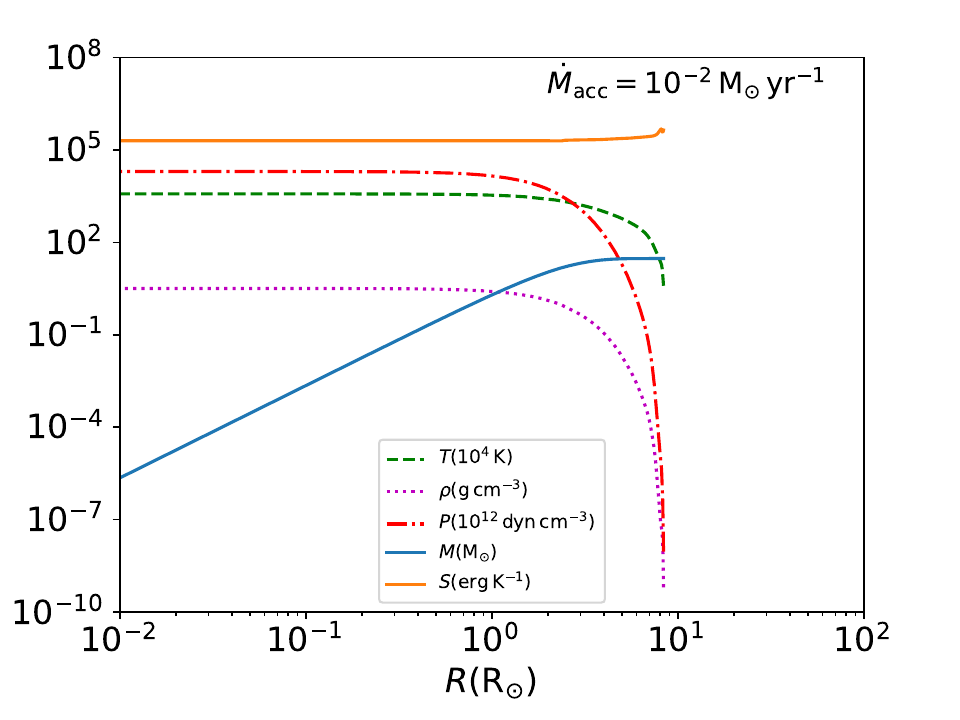}
    
    &
     \hspace{-1.2cm} 
    \includegraphics[width=.52\linewidth]{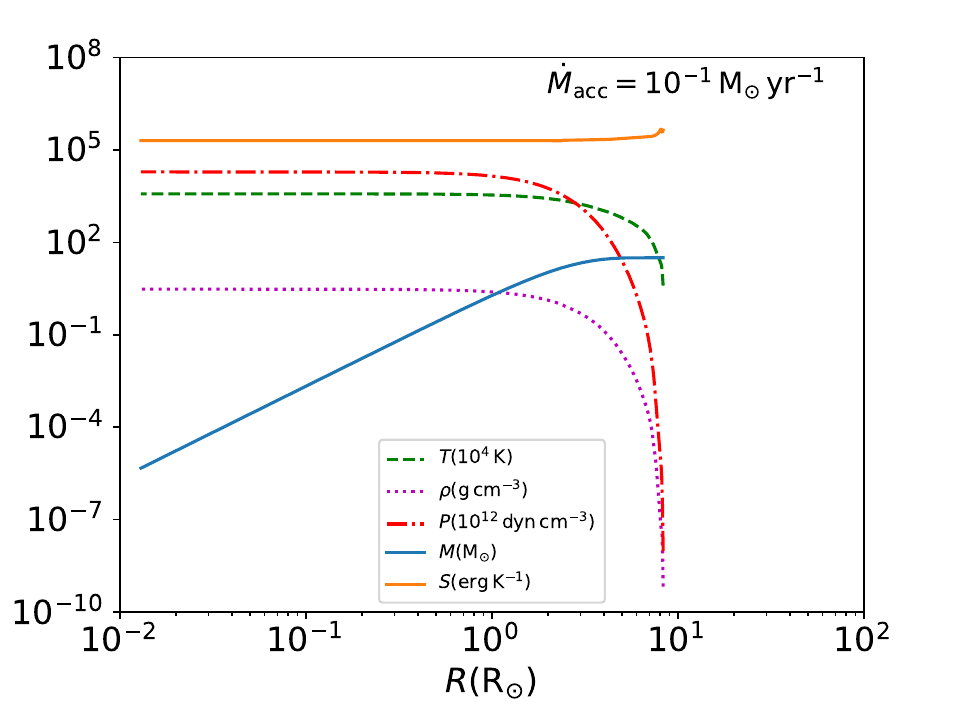}
    \\
    \small (c) & (d)
  \end{tabular}
  \caption{The stellar properties, including temperature, density, pressure, mass, and entropy of a $ \rm 30~M_{\odot}$ star for four different accretion rates, at points O, and B' (see Figure \ref{H_all}). Panel (a), panel (b), panel (c), and panel (d) correspond to the accretion rates $\rm 10^{-4}$, $\rm 10^{-3}$, $\rm 10^{-2}$, and  $\rm 10^{-1}~M_{\odot}~\rm yr^{-1}$ respectively. }
  \label{pointD}
\end{figure*}

Figure \ref{H_all} panel (d) displays the evolutionary track for an accretion rate of $10^{-1}~\rm M_{\odot }~\rm yr^{-1}$, which shows fluctuating luminosity in the initial days.
In this scenario, luminosity rises rapidly in the initial days from point B to X ($ \simeq 2.3~\rm days$, similar to the case with $ \rm \dot{M}_{acc} $= $\rm 10^{-2}~M_{\odot}~\rm yr^{-1}$). The star experiences a high accretion rate and does not maintain the thermal equilibrium. After that, the luminosity began to decrease (track from point X to Y, $ \simeq 46~\rm days $)  but then it increased again (track from point Y to C) and became constant. This fluctuation in luminosity is observed only at the high accretion rate i.e., $\rm 10^{-1}~M_{\odot}~\rm yr^{-1}$ during the accretion phase (see Section \ref{3.2} below). During this time, the star also inflates and moves towards the cooler side of the HR diagram. After reaching point C, it also enters the recovery phase and evolves towards the hotter side of the HR diagram. The track from point C to B' and B' to D represents the recovery phase in Figure \ref{H_all} panel (d). During the recovery phase, the star exhibits behaviour similar to that observed for the case with $\rm 10^{-2}~M_{\odot}~\rm yr^{-1}$. For these higher accretion rates the estimated time from point C to D is $\simeq 3.86$ Myrs.

We find that the star responds differently to lower and higher accretion rates. At higher accretion rates, the star moves towards the cooler side of the HR diagram during the accretion phase, while at lower accretion rates, it remains on the hotter side. In the recovery phase, the star with a lower accretion rate experiences radial inflation (as shown by the track from point B to D in Figure \ref{HR0}), whereas the star with a higher accretion rate shifts to the hotter side of the HR diagram and eventually follows a similar path to that of the lower accretion rate star. The stellar parameters for each accretion rate after 20 yrs of accretion (point C) is detailed in Table \ref{T2}.

\subsection{The cause and effect of the luminosity fluctuation in high mass accretion rate}
\label{3.2}
   To analyze the luminosity fluctuation at a higher accretion rate of $1\rm 0^{-1}~M_{\odot}~\rm yr^{-1}$ (see Figure \ref{H_all} panel (d) in the track from point B to Y), we perform simulations with four different accretion rates ranging between $ \rm 10^{-2}$ to  $\rm 10^{-1}~M_{\odot}~\rm yr^{-1}$. These rates are $\rm 2\times 10^{-2}$, $\rm 4 \times10^{-2}$, $\rm 6 \times10^{-2}$, and $\rm 8 \times 10^{-2} ~M_{\odot}~\rm yr^{-1}$. This analysis aims to identify the threshold accretion rates at which luminosity fluctuations occur and to determine the underlying cause of these variations.

   Figure \ref{L} displays the luminosity tracks for the six distinct accretion rates $\rm 10^{-1}$, $\rm 8\times 10^{-2}$, $\rm 6 \times 10^{-2}$, $\rm 4 \times 10^{-2}$, $\rm 2 \times 10^{-2}$, and $\rm 10^{-2}~M_{\odot}~\rm yr^{-1}$ over a period of 20 $\rm yrs$. As seen in Figure \ref{L}, luminosity for the accretion rates of $  \leq \rm 6 \times 10^{-2}~M_{\odot}~\rm yr^{-1}$ rises sharply in the initial days and become constant throughout the accretion phase. In contrast,  the luminosity for accretion rates of $> \rm 6 \times 10^{-2}~M_{\odot}~\rm yr^{-1}$ show a rapid initial rise, followed by a dip, and then another increase. This luminosity fluctuation appears around accretion rates of $ \simeq \rm 8 \times 10^{-2}~M_{\odot}~yr^{-1}$ and become more dominant for accretion rates higher than $\rm \gtrsim 8 \times 10^{-2} ~M_{\odot}~\rm yr^{-1}$ . Thus the rapid increase in the luminosity, followed by a later drop, is driven by changes occurring at the stellar surface.

Possible causes include the winds or compression of the outer layers as the material is added at each time step during the accretion phase. Wind mechanics cause the material to be blown off the outer layers, reducing the luminosity due to mass loss. However, in our simulations, these winds are insufficient to remove material from the star’s outer layers.  As shown in Figure \ref{all} panel (a), the Eddington parameter does not exceed $1$ during accretion. Additionally, Figure \ref{all} panel (b) confirms this, as the star’s mass steadily increases throughout the accretion phase without any discontinuities or drops in the mass value on the track. Also, Figure \ref{all} panels (c) and (d) show the changes in gravitational energy ($E_{\rm grav}$) and entropy ($S$) of the star's outer layers during the accretion phase. Using equation \ref{6}, we find that the second term on the right side represents $E_{\rm grav}$ loss due to compression in the outer layers, resulting from the accretion of material in the star's outer layers. As the material is added, $E_{\rm grav}$ initially increases; however, compression in the outer layers also causes some gravitational energy to be lost as given in equation \ref{6}. This results in a steady gravitational energy level in the outer layer during certain time steps of the accretion process, seen in Figure \ref{all} panel (c) as a constant line in the initial days. The substantial amount of added material and high temperature in the outer layers amplify this compression effect. During this period, the added $E_{\rm grav}$ does not contribute to outward luminosity, resulting in a drop in the luminosity tracks for the accretion rates higher than $\rm \gtrsim 8 \times 10^{-2} ~M_{\odot}~\rm yr^{-1}$. The threshold accretion rate where compression dominates in the outer layers is $ \dot M_{\rm acc}$~$ \simeq$~$\rm 8 \times 10^{-2}~M_{\odot}~\rm yr^{-1}$. Over time, as temperature decreases due to the added layers, the compression effect decreases, allowing luminosity to rise again, as illustrated in Figure \ref{H_all} panel (d).

\subsection{The evolution of the surface properties during the accretion}
Figure \ref{all}, panel (c) shows the variation in the $E_{\rm grav}$ during the accretion phase for four distinct accretion rates i.e., $\rm 10^{-4}$,  $\rm 10^{-3}$, $\rm 10^{-2}$, and  $\rm 10^{-1}~M_{\odot}~\rm yr^{-1}$ respectively. The change in the $E_{\rm grav}$ is more dominant for high accretion rates compared to lower accretion rates, due to the higher amount of material being added to the outer layers of the star. Table \ref{T2} presents the changes in $\Delta E_{\rm grav}$ for all accretion rates ($\dot{M}_{\rm acc}$) during the accretion phase, i.e., at t = 0 $ \rm yrs$ (point B) and t = 20 $ \rm yrs$ (point C). The added $E_{\rm grav}$ during the accretion phase follows the standard form given in equation \ref{5}. 
Similarly, panel (d) of Figure \ref{all} displays the variation in the $S$ resulting from the additional material that is added in the star's outer layers during the accretion phase for all values of $\dot{M}_{\rm acc}$.

Panels (a), (b), and (c) of Figure \ref{LTR} show how the luminosity ($L$), radius ($R$), and effective temperature ($T_{\rm eff}$) of the star's outer layers change during the accretion process. The star maintains a hydrostatic equilibrium structure through a balance between two opposing forces: gravity, which pulls inward and compresses the star, and the internal pressure maintained by nuclear fusion in the core, which pushes outward. However, this equilibrium can be disrupted when a significant amount of the mass is added to the star's outer layers, which causes the gravitational force to increase.
Using the control option in \textsc{mesa}, the material added during the accretion process has the same chemical composition and thermodynamic properties as the material that was present before the accretion onto the outer layers. The material accreted onto a star's outer layers indeed heats up and releases the potential energy. Then, it gains kinetic energy, which is subsequently converted into thermal energy. However, as the gravitational force increases due to accretion, the star is no longer in hydrostatic equilibrium, meaning that the virial theorem (which states that only half of the gravitational energy released is converted into thermal energy while the remainder is radiated) does not apply here. Here, we do not specify the fraction of thermal energy released during the accretion phase, as this aspect is not the focus of this study. Thus, this released thermal energy leads to an initial rapid increase in the star's luminosity, radius, and effective temperature, as shown in Figure \ref{LTR}, during the early days of the accretion process. Later on, these parameters stabilize and remain relatively constant over the following 20 yrs. For lower accretion rates, such as $\rm 10^{-3}$, and $\rm 10^{-4}~M_{\odot}~\rm yr^{-1}$, the changes in these stellar parameters are minimal, as the system remains in thermal equilibrium. In contrast, for higher accretion rates, like $\rm 10^{-2}$, and $\rm 10^{-1}~M_{\odot}~\rm yr^{-1}$, the changes are significant, as the system goes out from thermal equilibrium, resulting in rapid increases in $L$, $T_{\rm eff}$, and $R$. In the high accretion rate scenario, the added mass is significantly higher than in the lower accretion rate cases, leading to a dominance of the outer layers. As a result, $T_{\rm eff}$ begins to decrease while the radius ($R$) and luminosity ($L$) remain relatively constant. Thus the increased values of these stellar parameters are entirely dependent on the accretion rate.

\begin{table*}
\centering
    \begin{tabular}{l c c c c c}
    \hline \hline
    $\Dot{M}_{\rm acc} (\rm M_{\odot}~\rm yr^{-1})$ & No Accretion &  $ 10^{-4} $ &  $ 10^{-3}$ &  $ 10^{-2} $  & $ 10^{-1} $ \\
    \hline  log $ L_{\rm C}$ ~($ \rm L_{\odot}$) & 5.12 & 5.13 &  5.19 & 5.82 & 5.83   \\
      log $ T_{\rm eff, C}~ (\rm K)$ & 4.58 & 4.59 & 4.60 & 3.66 & 3.54    \\
      log $ R_{\rm C}$~($ \rm R_{\odot}$) & 0.91 & 0.91 & 0.92  & 3.10 & 3.33\\
      log $g_{\rm C}$ ~$( \rm cm~s^{-2})$ & 4.09 & 4.08 & 4.06 &  -0.28 & -0.73 \\
      $M_{\rm B}~(\rm M_{\odot})$ & 29.8577 &   29.8577  & 29.8577  & 29.8577  & 29.8577   \\
      $ \Delta M_{\rm added} ~(\rm M_{\odot})$ & 0 & 0.0020 & 0.020 & 0.20 & 2 \\
       $ M_{\rm C}~(\rm M_{\odot})$ & 29.8577 & 29.8597 & 29.8777 & 30.0577 & 31.8577  \\
      
      $ E_{\rm grav, B} ~(10^{50}~\rm erg)$ & $-6.14$ & $-6.14$ & $-6.14$ & $-6.14$ & $-6.14$ \\
      $ E_{\rm grav, C}~(10^{50}~\rm erg)$ & $-6.14$ & $ -6.14 $ &   $ -6.15$ &$ -6.17 $ & $ -6.18$ \\
      $ \Delta E_{\rm grav}~(10^{50}~\rm erg)$ & $ 3.4 \times 10^{-6} $ &  $ 8.9 \times 10^{-4} $ &   $ 8.9 \times 10^{-3} $ &  $ 0.03 $ & $ 0.04 $ \\ 
   
    $ E_{\rm int, B}~(10^{50}~\rm erg)$ & $ 37.0 $  & $ 37.0 $  & $ 37.0 $  & $ 37.0 $  & $ 37.0 $ \\
        $ E_{\rm int, C}~(10^{50}~\rm erg)$ & $ 3.70 $ & $ 3.70 $ & $ 3.71 $ & $ 3.72 $ & $ 3.73 $\\

     $ \Delta E_{\rm int}~(10^{50}~\rm erg)$ & $ 0.6 \times 10^{-5} $  & $ 3.3$ & $ 3.3 $ & $ 3.3 $& $ 3.4 $  \\

       $E_{\rm total, B}(10^{50}~\rm erg) $ & $ -2.43 $ & $ -2.43 $ &$ -2.43 $ &$ -2.43$ & $ -2.43 $  \\
       $E_{\rm total, C}(10^{50}~\rm erg)$ & $-2.43 $ & $ -2.44  $ & $ -2.44$ &  $-2.45 $ & $-2.45  $\\

        $ \Delta E_{\rm total}(10^{50}~\rm erg)$ & $ 2.7 \times 10^{-6} $ & $ 3.1 \times 10^{-4}  $ & $ 3.1 \times 10^{-3} $ &  $ 0.03  $ &  $ 0.03  $\\
         $S_{\rm B}(10^{36}~\rm erg~K^{-1})$ & $1.47 $ & $1.47 $ & $1.47 $ & $1.47 $ & $1.47$ \\
          $ S_{\rm C}(10^{36}~\rm erg~K^{-1})$ & $1.47 $ & $1.47 $ & $ 1.48$ & $ 1.49  $ & $ 1.67$\\
          
           $ \Delta S_{\rm total}(10^{36}~\rm erg~K^{-1})$ & $ 1.1 \times 10^{-6} $ & $1.2 \times 10^{-4}$ & $1.2 \times 10^{-3} $ & $ 0.02  $ & $ 0.20 $\\
      $ \Delta t ~( \rm s)$ & 500 & 500 & 500 & 500 & 500 \\
     log $ \Gamma_{\rm C}$ & -0.32 & -0.31 & -0.26 & -2.10 & -2.32\\
      log $  \tau_{\rm C}$ & -0.17 & -0.17 & -0.17 & -0.17 & -0.17  \\
      $E_{\rm err, cons,C}(\times 10^{-13})$ & $ 4.99 $ & $ 4.87 $ &  $ 4.90 $ &  $ 4.96 $ & $ 4.93 $\\  
      \hline\hline
       \end{tabular}
       \caption{ Stellar parameters of the $ \rm 30~M_{\odot}$ star at the initial stage (t = 0 $ \rm yrs$, point B), and final stage (t = 20 $\rm yrs$, point C) during the mass accretion process are shown above. The rows are, from top to bottom: surface luminosity at the final stage of the accretion phase ($ L_{\rm C}$), surface temperature at the final stage of the accretion phase ($ T_{\rm eff, C}$), stellar radii at the final stage of the accretion phase ($ R_{\rm C}$), surface gravity at the final stage of the accretion phase ($ g_{\rm C}$), mass of the star at the initial stage of the accretion phase ($ M_{\rm B}$), mass added over the 20 $ \rm yrs$ ($ \Delta M_{\rm added}$), the mass of the star at the final stage of the accretion phase ( $M_{\rm C}$), gravitational energy at t = 0 $\rm  yrs$ ($ E_{\rm grav, B}$), gravitational energy at t = 20 $\rm yrs$ ($ E_{\rm grav, C}$), added gravitational energy over 20 $ \rm yrs$ ($ \Delta E_{\rm grav}$), internal energy at t = 0 $ \rm yrs$ ($ E_{\rm int, B}$), internal energy at t = 20 $\rm  yrs$ ($ E_{\rm int, C}$), change in internal energy over 20 $ \rm yrs$ $ \Delta E_{\rm int}$, total energy at t = 0 $ \rm yrs$ ($ E_{\rm total, B}$), total energy at t = 20 $ \rm yrs$ ($ E_{\rm total, C})$, change in total energy over 20 $ \rm yrs$ ($ \Delta E_{\rm total}$), total entropy at t = 0 $ \rm yrs$ ($ S_{\rm B}$), total entropy at t= 20 $\rm  yrs$ ($S_{\rm C}$), added entropy over 20 $ \rm yrs$ ($\Delta S_{\rm total}$), time step for the accretion ($ \Delta t$), Eddignton parameter at the final stage of the accretion phase ($ \Gamma_{\rm C}$), surface optical depth at the final stage of the accretion phase ($ \tau_{\rm C}$), and relative error in energy conservation in the evolution of $ \rm 30~M_{\odot}$($ E_{\rm err, cons, C}$).}
       \label{T2}
      \end{table*}

\subsection{Recovery}
The recovery period of the primary star following a GE is typically long, often lasting years or even centuries \citep[e.g.,][]{2009NewA...14..539H, Davidson_2012, 2016ApJ...817...66K}. However, the post-accretion recovery phase of a companion star has not been studied in detail. Therefore, we simulate the companion star's evolution after the accretion phase to analyze its recovery process.

Panel (a) of Figure \ref{H_all} shows the evolutionary track for an accretion rate of $\rm 10^{-4}~M_{\odot}~\rm yr^{-1}$, including both the accretion and recovery phases.  As mentioned earlier in Section \ref{3.2}, the $T$ is high at point C ($\log T \simeq 4.60~\rm K$ ), as the star enters the recovery phase, it experiences Dutch stellar wind mass loss which causes the decrease in luminosity from point C to O. In this scenario, the mass loss rates range from $ \rm 10^{-6}~M_{\odot}~\rm yr^{-1}$ to $ \rm 10^{-5}~M_{\odot}~\rm yr^{-1}$. 
After that, the star undergoes radial inflation (track from point O to D) and evolves towards the cooler side of the HR diagram. As a result, $T$ begins to drop. During this evolution, the star does not remain hot enough to drive stellar winds, leading to a minimal mass loss rate. It enters the helium-burning phase and undergoes radial expansion. The total mass lost during the recovery phase is $\Delta M = 0.7610~\rm M_{\odot}$. Similarly, panel (b) of Figure \ref{H_all} shows the evolutionary track for an accretion rate of $\rm 10^{-3}~M_{\odot}~\rm yr^{-1}$. The track from point C to O and from O to D represents the recovery phase of the star, during which the total mass lost amounts to $\Delta M = 0.7643~\rm M_{\odot}$.
The mass lost during the recovery phase is greater for an accretion rate of $\rm 10^{-3}~M_{\odot}\rm yr^{-1}$ compared to $\rm 10^{-4}~M_{\odot}\rm yr^{-1}$. This is because, at point C, the $\rm 10^{-3}~M_{\odot}\rm yr^{-1}$ accretion rate results in higher temperature and luminosity, leading to increased mass loss through stellar winds.

Furthermore, panels (c) and (d) of Figure \ref{H_all} show the evolutionary tracks for the higher accretion rates of $\rm 10^{-2}$, and  $\rm 10^{-1}~M_{\odot}~\rm yr^{-1}$, respectively. In both cases, point C represents the final stage of the accretion phase. After this point, as the star transitions into the recovery phase, it moves from the cooler side to the hotter side of the HR diagram and continues to evolve. This contraction back toward the ZAMS occurs when the accreting star becomes sufficiently luminous and large to dissipate the excess gravitational energy of the envelope material as mentioned in \citet{2024ApJ...966L...7L}. Later on, the star reaches the final stage of recovery, point D, after crossing the TAMS. The total mass lost during the recovery phase is $0.7748~\rm M_{\odot}$ and $0.912~\rm M_{\odot}$ for accretion rates of $\rm 10^{-2}$ and $\rm 10^{-1}~M_{\odot}\rm yr^{-1}$, respectively. At points C, B', and D, the star's ages are $1.159452$, $1.160186$, and $5.020271~\rm Myrs$, for the accretion rate $\rm 10^{-2}~M_{\odot}~\rm yr^{-1}$. This indicates that the recovery phase (from point C to B') takes $\approx 734~\rm yrs$ for higher $\dot{M}_{\rm acc}$. Thus during the recovery phase, the star undergoes a process of readjustment, where it reforms its structure and composition to a new equilibrium state.

Figure \ref{pointC} shows the stellar properties, including $T$, $\rho$, $P$, $S$, and $M$, at the final stages of the accretion phase, i.e., point C (see Figure \ref{H_all}), for all values of $\dot M_{\rm acc}$. Panels (a) and (b) present the stellar properties for the lower accretion rates, $\rm 10^{-4}$ and $\rm 10^{-3}~M_{\odot}~ yr^{-1}$, respectively. While panels (c) and (d) present the stellar properties for the higher accretion rates, $\rm 10^{-2}$ and $\rm 10^{-1}~M_{\odot } ~yr^{-1}$, respectively. The star remains on the hotter side of the HR diagram for the lower accretion rates, with a radius of $ \approx \rm 10~R_{\odot}$ at the end of the accretion phase. In contrast, at higher accretion rates, the star becomes cooler, with its radius expanding to nearly $\rm 10^{3}~R_{\odot}$, as shown in panels (c) and (d). As a result, $T$, $\rho$, $P$, and $S$ are higher in the outer regions for lower accretion rates than those for higher accretion rates. The star at the end of the accretion point is not one unit but rather composed of the original star and on top of it resides a layer that arrived from the accreted material and is separated in all its properties. Figure \ref{pointD} illustrates the stellar properties, including $T$, $\rho$, $P$, $S$, and $M$, at points O and B'. Panels (a) and (b) depict the stellar properties for lower accretion rates of $\rm 10^{-4}$ and $\rm 10^{-3}~M_{\odot}\rm yr^{-1}$, respectively. Panels (c) and (d) show the stellar properties for higher accretion rates of $\rm 10^{-2}$ and $\rm 10^{-1}~M_{\odot}\rm yr^{-1}$, respectively. Both points O and B' occur during the recovery phase, close to point B (as shown in Figure \ref{HR0}), where the accretion process is initiated. Despite corresponding to different accretion rates, points O and B' exhibit similar stellar properties.

\section{Discussion}
\label{4}

\begin{figure*}
  \includegraphics[trim= 0.0cm 0.0cm 0.4cm 0.0cm,clip=true,width=0.85\textwidth]{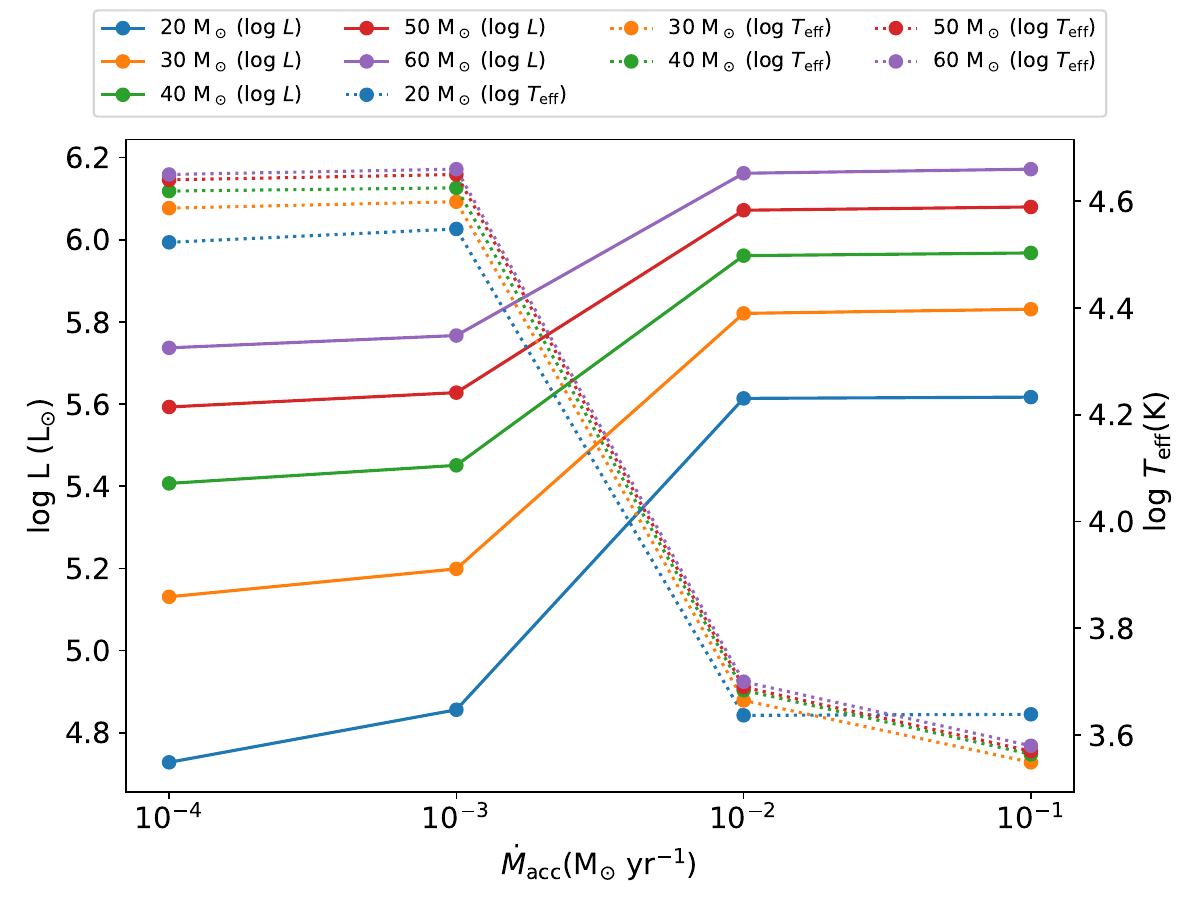} 
  \centering
  
\caption{The grid of massive stars with initial masses ranging from 20 to 60 $\rm M_{\odot}$ for which we simulated accretion. The accretion process for each star is initiated when its temperature approaches $\log T \approx 4.58~\rm K$. Four different accretion rates, ranging from $10^{-4}$ to $10^{-1}~\rm M_{\odot}~yr^{-1}$, are simulated for each of the 5 stars. The luminosity and temperature of each star at the end of the accretion phase are presented as the output of the simulation on the left and right side of the y-scale respectively.} 
\label{hr}
\end{figure*}

Our simulations follow the evolutionary tracks and changes in stellar parameters during the accretion phase. We identify the rate at which accretion influences the star's outer layers, particularly through compression observed in these layers.
We aim to understand how the incoming material, added to the star's outer layers through accretion, interacts with the existing envelope material. This interaction alters the stellar properties of the outer layers, including luminosity, temperature, and radius. We also investigate how the accretion phase impacts the evolutionary tracks of the star and assess whether the added material introduces any additional effects.

Other than our fiducial case presented in the previous sections, where the mass of the companion is 30 $\rm M_{\odot}$, we simulated four more accreting companions with masses of 20, 40, 50 to 60 $\rm M_{\odot}$. All these star are assumed to be in a binary system with the primary erupting at a close distance such that the accretion rate is very high, ranging from $10^{-4}$ to $10^{-1}~\rm M_{\odot}~yr^{-1}$ (in total we have 20 simulations for five stellar masses and four accretion rates).
Figure \ref{hr} shows for our grid of simulations the luminosity and temperature of each star at the end of the accretion phase.

Although it may seem that a $60\, \rm M_{\odot}$  companion for an $\sim 100\, \rm M_{\odot}$ erupting LBV is very massive, such a massive companion has already been suggested for $\eta$ Car by a few groups and is established both theoretically and observationally \citep{2010ApJ...710..729M,2010ApJ...723..602K,2023MNRAS.519.5882S}.
In the case of $\eta$ Car the orbit is $\sim 5.53$ yrs at an eccentricity of $0.9$ \citep[e.g.,][]{1997ARA&A..35....1D, 2012ASSL..384.....D} and facilitates such massive stars with very close periastron approach of less than twice the primary radius. Less eccentric systems will be able to host such massive stars at shorter orbital periods. Our model particularly describes massive binary systems with masses ranging from $100-200 \, \rm M_{\odot}$ with a period of a few months to a few years. Therefore, we consider such systems possible.

Systems with less massive companions might also exist, but companions with masses $ \leq 20 \rm \, M_{\odot}$ are not expected for a $\sim 100\, \rm M_{\odot}$ LBV as mass transfer at earlier stages would bring up the mass ratio (see recent work by \citealt{2025A&A...695A.117N} and references therein) and would likely result $q \geq 0.2$ by the time eruptions are expected to occur.

\citet{2010ApJ...723..602K} studied the companion star during the periastron passage that triggered the 19$^{\rm th}$ century $\eta$ Car eruptions. They found that the periastron passage resulted in a significant increase in the luminosity of the companion star, by an order of magnitude, for an accretion rate of $\rm \simeq 0.1-0.2~M_{\odot}~yr^{-1}$. Our study examined the significant impact of mass accretion on a star's evolution, particularly in its outer regions, where even small changes in accretion rates can lead to substantial alterations in surface properties. We find that for high accretion rates ($\rm  10^{-2}$ and $\rm 10^{-1}~M_{\odot}~yr^{-1}$), the increase in luminosity is $\Delta \log L \simeq 0.72$ dex, which is of the same order of magnitude with \citet{2010ApJ...723..602K}. However, for lower accretion rates, the change in luminosity is much smaller, as expected.
In order for GEs to be powered by accretion of primary material onto the companion, the companion should not expand by a large amount. In our simulations we treated the case of constant accretion at high rates without free parameters and external effects, and did not impose a prescription that regulates the expansion. Thus, we obtained large expansion when the accretion rates were high.
Such a prescription that iterates between mass accretion and removal in order to limit the expansion was recently used by \citep[][]{Scolnic_2025}, and was able to keep the accretor at a small radius.

\section{Summary}
\label{5}

We examine how a massive companion star in a binary system reacts to the accretion of gas arriving at a high rate from a primary star undergoing a Giant Eruption. The mass accretion rate is influenced by factors such as the mass loss rate of the erupting star, the binary separation, and the radiation field of the accreting star, which can limit the amount of material that accretes. We used a simplified approach with a constant accretion rate and demonstrated how the star responds to mass accumulation in its outer layers, resulting in changes to its surface stellar parameters.

Using \textsc{mesa} we modelled the evolution of an accreting star. We created a grid of stars with masses ranging from 20 to 60 $\rm M_{\odot}$ and used four different accretion rates to observe the star's response to the accreting mass. Figure \ref{1} shows the evolution of these stars using \textsc{mesa}'s default settings, while Figure \ref{2} shows the diagram showing the star's response to the different accretion rates. Later in the article, we presented a detailed analysis of a 30 $\rm M_{\odot}$ star, focusing on its response to all the accretion rates as well as its recovery phase. Here Figure \ref{1} shows the evolution of 30 $\rm M_{\odot}$, as it evolves without any accretion from the ZAMS to the helium-burning phase. After the MS phase, the star undergoes stellar wind mass loss, passes the TAMS, and enters the helium-burning phase.  At this stage, the star’s mass is $ \rm 29.09~M_{\odot}$. Figure \ref{2} shows the output of the accretion process on a $\rm 30~M_{\odot}$ star, with simulations run for 20 yrs using four different accretion rates applied to the star's outer layers, starting at point B in Figure \ref{1}. Our work reveals that the star responds differently to lower and higher accretion rates, as shown in Figure \ref{H_all}. At high accretion rates ($\rm 10^{-2}$, $\rm 10^{-1} \, M_{\odot}\rm\, yr^{-1}$), the star becomes cooler during the accretion phase. During the recovery phase, it contracts toward the ZAMS and continues its evolution until point D. At lower accretion rates ($\rm 10^{-3}$ and $\rm 10^{-4}~M_{\odot}\,\rm yr^{-1}$), the star stays on the hotter side of the HR diagram during the accretion phase. During the recovery phase, due to its high temperature, it undergoes stellar wind mass loss shortly after accretion, expelling material and leading to a drop in luminosity. After that, it undergoes radial inflation and reaches point D. Our results indicate that during the initial days of accretion, both luminosity ($L$) and radius ($R$) increase rapidly for all values of $\dot{M}_{\rm acc}$. The changes are more significant for higher accretion rates, while they are minimal for lower accretion rates as shown in Figure \ref{LTR}. This phenomenon occurs as gravitational energy from the added material is converted into thermal energy, raising the $L$ and $R$ of the star’s outer layers. Thus the surface luminosity is driven up almost immediately and adiabatically towards the accretion luminosity. Table \ref{T2} presents the values of all stellar parameters and all kinds of energy values during the accretion process. However, not all of the energy from the added material is radiated outward. For higher accretion rates, about 46\% of the additional energy contributes to an increase in luminosity, whereas for lower accretion rates, only $\simeq$ 23\% of the total energy is radiated outward, and increases the luminosity.
However, as the star continues to evolve, its  $L$ and $R$ reach a constant value, while the $T_{\rm eff}$ decreases due to the dominance of the outer layers in the case of high accretion rates. In contrast, $T_{\rm eff}$ changes are minimal for lower accretion rates. Furthermore, our simulation showed that at high accretion rates, such as  $\rm 10^{-1}$ and $\rm 8 \times 10^{-2}~M_{\odot}~\rm yr^{-1}$, the compression caused by the added material in the outer layers also becomes significant, leading to fluctuations in luminosity during the initial days of the accretion phase (see Figure \ref{L}). This work helped us to identify the accretion rates at which compression in the star's outer layers becomes dominant.

Our model does not account for any stellar rotation, tides and magnetic effects. Instead, we focus on studying the impact of stellar inflation in isolation. In future work, we will show how stars at different metallicities respond to high accretion rates. Given these uncertainties and the large parameter space, as well as the numerical challenges, many more simulations are required to improve our understanding of how a star reacts to accretion.

\vspace{0.5cm}
We thank an anonymous referee for very helpful comments. BM acknowledges support from the AGASS center at Ariel University.
We acknowledge the Ariel HPC Center at Ariel University for providing computing resources that have contributed to the research results reported in this paper.
The MESA inlists and input files to reproduce our simulations and associated data products are available on Zenodo (\url{DOI:10.5281/zenodo.15050466}).

\vspace{0.5cm}

\section*{APPENDIX: Treatment of Mass Accretion}

\label{treatment}

We use \textsc{mesa} control default option `mass change' to apply the accretion onto the outer layers of the star.
Simulating accretion in \textsc{mesa} requires a special approach that handles the interaction of the arriving material with the outer layers of the star, and in every step gravitational energy is added to the gas in the outermost cell.
The specific rate of change in gravitational energy due to contraction or expansion of the outer layers is \citep[e.g.,][]{1990sse..book.....K, 2009itss.book.....P}
\begin{equation}
     \\  \epsilon_{\rm grav} = -T   \left( \frac {\partial S} {\partial  t }\right)_{ m},
 \label{eq1}
\end{equation}
where $S$ is the entropy.
\textsc{mesa} computes the gravitational energy by expressing it in the form of local thermodynamic independent variables e.g., temperature ($T$) and density ($\rho$). Thus, rewriting the equation \ref{eq1} \citep[e.g.,][]{1990sse..book.....K}
\begin{equation}
\begin{split}
\epsilon_{\rm grav} = -C_{p}T \Bigg[ 
    \big(1 - \nabla_{\rm ad} \chi_{T} \big) 
    \left(\frac{\partial \ln T}{\partial t}\right)_{m} \\
    - \nabla_{\rm ad} \chi_{\rho} 
    \left(\frac{\partial \ln \rho}{\partial t}\right)_{m} 
\Bigg],
\label{eq2}
\end{split}
\end{equation}

Here, all the thermodynamic quantities have their usual symbols, and the thermodynamic derivatives are $\nabla_{\rm ad} $ = $\left(\frac{\partial \ln T}{\partial \ln P}\right)_{s}$, $ \chi_{T} $= $\left(\frac{\partial P}{\partial T}\right)_{\rho}$, and $\chi_{\rho} $ = $\left(\frac{\partial P}{\partial \rho}\right)_{T}$.
 \textsc{mesa} calculates these time derivatives using the quantities provided by the equation of state (EOS) \citep{2002ApJ...576.1064R} to determine the $ \epsilon_{\rm grav}$.  Using the radiation$+$gas equation of state $ P = P_{\rm gas} + P_{\rm rad}$, 
 \textsc{mesa} treats $P_{\rm gas} $ as the basic variable instead of $ \rho$ \citep[e.g.,][]{2011ApJS..192....3P, 2013ApJS..208....4P}, and
 equation \ref{eq2} becomes
\begin{equation}
\begin{split}
    \epsilon_{\rm grav} = &-C_{p}T\left[\left(1-4\nabla_{\rm ad}\frac{P_{\rm rad}}{P}\right)\left(\frac {\partial \ln T}{\partial t}\right)_{m}\right]\\
    & + \nabla_{\rm ad} \frac{P_{\rm gas}}{P} \left(\frac{\partial \ln P_{\rm gas}}{\partial t}\right)_{m} .
    \label{eq3}
    \end{split}
\end{equation}

The accretion process increases the mass in the outer layers of the star and leads to changes in the entropy of the system. So it is essential to determine the thermal state of the recently accreted material of mass $m$ in its outermost layers to the present mass of the star $M$. In this context, we need two relevant timescales: the local accretion time $\tau_{\rm acc} \simeq \frac{(M-m)}{\dot{M}}$, and the thermal time $\tau_{\rm th} = \frac{(M-m)C_{p}T}{L} $. Near the surface, $ L \gg C_{P}T{\dot{M}}$, thus $ \tau_{\rm th} < \tau_{\rm acc}$ holds, allowing the outer layers sufficient time to return to the thermal equilibrium configuration \citep[e.g.,][]{1982ApJ...253..798N, 2004ApJ...600..390T}. 
Thus, we can use the estimate that the accreting material has the same entropy as the photosphere. This implies that the thermal state of the incoming material becomes irrelevant. In this scenario, the accretion shock's (or boundary layer's) luminosity is \citep{2024ApJ...966L...7L}:
\begin{equation}
     L_{\rm acc} = \frac{Gm{\dot{M}_{\rm acc}}}{R} .
     \label{eqL}
\end{equation}

It radiates outward and does not impact the material's entropy as it integrates into the hydrostatically adjusted star. While the transit of $ L$ determines the added material's entropy \citep[e.g.,][]{1982ApJ...253..798N, 2004ApJ...600..390T, 2013ApJS..208....4P}. The timescale hierarchy in accretion suggests that the outer regions evolve in a nearly homologous manner. As a result, the thermal profile, such as the variation of temperature ($T$) with pressure ($P$) or density $ \rho$, in the outer layer remains relatively constant over time. This occurs despite the compression of fluid elements, which added to the outermost cell higher pressure, causing an increase in the temperature of these elements. This motivates reformulating the equation \ref{eq1} of $ \epsilon_{\rm grav}$ \citep[see e.g.,][]{1982ApJ...253..798N, 2004ApJ...600..390T, 2011ApJ...732...20K, 2013ApJS..208....4P}
\begin{equation}
    \epsilon_{\rm grav} = -T  \left(\frac {\partial S } {\partial  t }\right)_{q} + \Delta \epsilon_{\rm grav},
    \label{L2}
\end{equation}
where $ \Delta \epsilon_{\rm grav}$ represents additional gravitational energy due to compression in the outer layers. Rewriting equation \ref{L2} we get
\begin{equation}
      \epsilon_{\rm grav} = -T  \left(\frac {\partial S } {\partial  t }\right)_{q}  + T ~\frac{\partial \ln M(T)}{\partial t} \left(\frac{\partial S}{\partial \ln q}\right)_{t},
      \label{6}
\end{equation}
where $q$ is the fractional mass coordinate ($q=m/M$) and $2^{\rm nd}$ term on the right-hand side of the equation \ref{6} represents the local energy loss that occurs when the fluid element is compressed due to higher pressure. Thus rewriting equation \ref{eq2} gives
\begin{equation}
\begin{split}
    \epsilon_{\rm grav} &= -C_{p}T\left[\left(1-\nabla_{\rm ad}\chi_{T}\right)\left(\frac {\partial \ln T}{\partial t}\right)_{m} - \nabla_{\rm ad} \chi_{\rho}\left(\frac{\partial \ln \rho}{\partial t}\right)_{m} \right] \\ 
    &+ \frac{C_{P}TGm{\Dot{M}}}{4 \pi r^{4}P} \left( \nabla_{\rm ad} - \nabla_{T}\right),
\end{split}
    \label{7}
\end{equation}
where $ \nabla_{T} = \frac{\partial \ln T}{\partial \ln P}$. Thus, equation \ref{5} represents the total $\epsilon_{\rm grav}$ of the system during the mass accumulation process .\\

\vspace{0.5cm}

\bibliography{Ref}
\bibliographystyle{mnras}

\end{document}